\documentclass[11pt,a4paper]{article}
\usepackage{jcappub}

\usepackage{graphicx}
\usepackage{epstopdf}
\usepackage{mathrsfs}
\usepackage{amssymb}
\usepackage[title]{appendix}
\usepackage{amsmath}
\usepackage{color}
\usepackage{bm}
\usepackage{color,xcolor}

\newcommand{\pri}{\prime}

\newcommand{\vu}{\mathbf u}

\newcommand{\vphi}{\mathbf \phi}

\newcommand{\vx}{\mathbf x}
\newcommand{\vk}{\mathbf k}
\newcommand{\vq}{\mathbf q}
\newcommand{\vp}{\mathbf p}

\newcommand{\mP}{\mathcal P}
\newcommand{\mT}{\mathcal T}
\newcommand{\mF}{\mathcal F}

\newcommand{\mH}{\mathcal H}

\newcommand{\mL}{\mathcal L}

\newcommand{\mS}{\mathcal S}
\newcommand{\mC}{\mathcal C}

\newcommand{\mD}{\mathcal{D}}

\newcommand{\vPsi}{\bm \Psi}
\newcommand{\vGamma}{\bm \Gamma}
\newcommand{\vgamma}{\bm \gamma}
\newcommand{\vpsi}{\bm \psi}
\newcommand{\vchi}{\bm \chi}

\newcommand{\vX}{\bm X}
\newcommand{\vY}{\bm Y}

\newcommand{\vpi}{\bm \pi}
\newcommand{\vomega}{\bm \omega}

\newcommand{\xiYinv}{\left(\xi^Y\right)^{-1}}
\newcommand{\tildeLambda}{\widetilde{\Lambda}}
\newcommand{\tildeGamma}{\widetilde{\Gamma}}

\newcommand{\ini}{0}

\newcommand{\vpsieff}{\vpsi_{{\rm eff}, \Lambda}}

\newcommand{\revision}{}
\newcommand{\revone}{}

\makeatletter

\newcommand{\rom}[1]{\expandafter\@slowromancap\romannumeral #1@}
\newcommand{\Rmnum}[1]{\textup{\uppercase\expandafter{\romannumeral#1}} }
\makeatother

\title{On the Statistical Evolution of the Large-scale Structure and the Effective Dynamics}

\author{Xin Wang}
\affiliation{School of Physics and Astronomy, Sun Yat-Sen University, No.2 Daxue Rd, 519082, Zhuhai, China}
 \emailAdd{wangxin35@mail.sysu.edu.cn}
\date{\today}

\label{firstpage}

\abstract{
In this paper, we study the statistical evolution of the large-scale structure (LSS), focusing on the joint probability distribution function (PDF) of the coarse-grained cosmic field and its role in constructing effective dynamics.
As the most comprehensive statistics, this PDF encodes all cosmological information of large-scale modes, therefore, could serve as the basis in the LSS modelling. Following the so-called PDF-based method from turbulence, we write down this PDF's evolution equation, which describes the probability conservation. 
We show that this conservation equation's characteristic curves follow the same PDF history and could be considered as an effective dynamics of the coarse-grained field. 
Unlike the EFT of LSS, which conceptually would work at both realization and statistics level, this effective dynamics is valid only statistically.  {\revision 
However, this `statistical equivalence' also provides valuable insight into scale interactions at the statistical level.
It also enables predicting a wide variety of statistics beyond the typical N-point polyspectra, including, e.g. topologies, density PDF and non-linear covariance matrices etc. 
Our formula expresses the small-scale effect as the ensemble average of their interactions conditional on the large-scale modes. This suggests an interesting way to measure effective terms directly from simulation.}
By applying the Gram-Charlier expansion, we demonstrate a different structure of these effective terms.  
This formalism is a natural framework for discussing the evolution of statistical properties of large-scale modes, and provides an alternative view for understanding the relationship between general effective dynamics and standard perturbation theory.
} 


\begin{document}
\maketitle

\section{Introduction}
The large-scale structure contains valuable information about our Universe, including the evolution history, composition, and primordial physics, all of which can be extracted from the statistical measurement of observed data. Consequently, the study of LSS has mainly been focusing on understanding the statistical properties of various cosmic fields  \cite{Peebles1980book,BCGS12review}. However, gravitational non-linearity renders the field non-Gaussian and complicates the theoretical calculation with the perturbation theory (PT) \cite{Peebles1980book,BCGS12review,GGRW86,JB94,MSS92}. 
The standard perturbation theory does not converge at non-linear scales after these quantities , e.g. the density contrast $\delta$, become non-perturbative \cite{CS06a}. 
As the large-scale surveys require higher and higher accuracy for describing the statistics of LSS, enormous efforts have been made to achieve this goal. 
In Eulerian space, the renormalized perturbation theory (RPT) \cite{CS06a,CS06b,CSB12,CS08BAO}  and its general extension, i.e. the Gamma expansion \cite{BCS12}, has been shown to obtain reasonable accuracy with a modest numerical cost \cite{BCS08,CSB12}.
In Lagrangian space, resummation technique have also been developed \cite{Matsubara08,OTM11}  and further improved to incorporate redshift distortion, clustering bias and non-Gaussian initial conditions \cite{Matsubara08bias,Matsubara12,WS12pert,Matsubara14}.
However, even with these sophisticated techniques, the agreement with simulation is still not satisfactory \cite{CWP09}. 
Moreover, despite valuable physical insights these techniques have provided, they are not perfect for calculating equal time correlation as they violate Galilean invariance, and result in only a partial resummation of large-scale modes, whose effect is cancelled in a more systematic treatment \cite{AP12,BGK13}.

Given these difficulties, another approach that utilizes the concept of the effective field theory (EFTofLSS) has been developed in recent years \cite{Baumann12,CHS12,H12,PZ13}. 
Various aspects have been discussed in the literature, e.g. in Eulerian space \cite{CFGS2lp14,CFGS14,BMZ15}, in Lagrangian space \cite{PSZ14},  with  ressumation \cite{SZ15}, bispectrum \cite{BMMP15,AFSS15}, biased tracer \cite{ABGZ14} etc.
Unlike the traditional perturbation theory, the EFT approach focuses on the coarse-grained field since, for many applications like the baryonic acoustic oscillation (BAO), we are only interested in the linear and quasi-linear regimes.  It systematically incorporates little-known small-scale information by introducing an effective stress-energy tensor with calibrated parameters. 
The hope is that by introducing these effective terms, one could capture highly non-linear mode couplings and the shell-crossing effects that previous perturbative calculations did not even try to address.

Generally speaking, most physical models could be considered `effective' to some extent, as lots of  dynamical degree-of-freedoms have to be neglected to reduce the problem's complexity.
For the large-scale structure, since we are only interested in the statistics of those cosmic fields,  constructing an effective theory of LSS would eventually come down to the fitting of some specific statistical measurements. 
For example, the excursion set theory of halo mass function \cite{excBond91,excZ07}, where we have dramatically simplified the much more complicated physical process to a  stochastic first-crossing problem only for recovering the statistics, i.e. the mass function.

As for the coarse-grained LSS field, the most comprehensive statistics one will ever need is the joint probability distribution function (PDF) of these large-scale modes. 
{\revision 
Of course, one has to reduce its dimension significantly for it to be observationally accessible. The examples include the first couple of moments, i.e. the power spectrum and bispectrum; PDF of local density, or some functions of these modes like Minkovski functional or other topological measurements.  }
Nevertheless, theoretically all cosmological information is encoded within. 
With those enormous number of degree-of-freedoms, the evolution history of this PDF provides only one constraint. Consequently, there could exist alternative dynamical systems that produce {\it exactly the same} PDF evolution. For statistical modelling of LSS, these alternative systems are indistinguishable from the original cosmic dynamics.  
Interestingly, one can write down the evolution equation of this PDF  at least formally. Such formalism stands at the core of the so-called {\it ``PDF-based method''} \citep{Pope1985} in turbulence. Similar equations can be found e.g. in \cite{Blas15,Blas16,WS16} in cosmological context  as well.  
{\revision Moreover, this framework describes the effective terms as the ensemble average of small-scale interactions conditional on large-scale modes. So instead of parametric fitting, this suggests an interesting way to directly measure the effective terms in simulation.}
Therefore, it provides a compelling perspective to constructing the effective theory in LSS.

{\revision
The prospect that we can construct a {\it ``statistically equivalent''} dynamics that recover the full joint PDF of the coarse-grained field is exciting.  
For one, it means that we could predict various statistics beyond N-point correlation functions, e.g. density PDF, topologies and covariance matrices etc.
For example, \cite{WS16} applied a similar method to study the evolution of a given Lagrangian fluid parcel. There, the effective parameter is described by the conditional average of the tidal tensor. It is not hard to see that, with further numerical measurements, the formula could grow to become an effective theory of local PDF of density and velocity gradient. 
Note that this is different from recent developments of count-in-cell PDF based on the large deviation principle \cite{Bernardeau_R_2016,Uhlemann_2017,Reimberg_2018},  which still assumes the spherical collapse model.   
\cite{Ivanov_2019} did incorporate aspherical contributions and EFT corrections, but the dynamics are still described traditionally.
}

Therefore, in this paper, we apply the PDF-based method to study the large-scale structure. {\revision As an initial investigation, the goal here is to first build the foundation of the framework, and then to understand its internal structure in more details.  }
In section 2, we first briefly introduce the core concept and its main conclusion of our approach. In section 3, we provide a detailed derivation of the PDF evolution of coarse-grained cosmic fields and then discuss its effective solution. In section 4, we discuss the various aspects in the perturbation theory and finally we conclude in section 5. 
Notably, we calculated the one-loop prediction of effective terms, explicitly verifying its recovery to the standard perturbation theory. These technical details are presented in Appendix.

\section{Brief Overview}
Before presenting the full derivation of our formalism, here we first provide a brief overview of our approach. At the sub-horizon scale, the dynamics of the structure formation is well described by the following fluid Poisson system \citep{BCGS12review}: 
\begin{eqnarray}
	\label{eqn:cdyn_eul}
	\partial_{\tau} \delta + \nabla_i  \left [ (1+\delta) u_i \right] &=& 0,  \nonumber \\
	\partial_{\tau} u_i + (u_j \nabla_j) u_i + \mH u_i &=& - \nabla_i \Phi - \pi_i,  \nonumber \\
	\nabla^2 \Phi &=& 4\pi G \bar{\rho} a^2 \delta, 
\end{eqnarray}
where $\delta=\rho/\bar{\rho}-1$ , $u_i$ is the peculiar velocity, $\mH(\tau) = d\ln a/d\tau$, and $\Phi$ is the gravitational potential. Here we also included the contribution $\pi_i =  (\nabla_j \rho \sigma^u_{ij} ) / \rho$ where $\sigma^u_{ij}$ is the stress tensor. 
Since the vorticity $\omega_i$ only arises after the shell-crossing, this system could be expressed in a nice compact form in Fourier space with the doublet definition $\psi_a (\vk) =\{  \delta(\vk), -\theta(\vk)/\mH \}$ with $\theta$ being the divergence of peculiar velocity \citep{BCGS12review}: 
\begin{eqnarray}
\label{eqn:fourier_dyn_full}
\hat{\mL}_{ab}  \psi_b (\vk)  
= \gamma_{abc}(\vk_1, \vk_2) \psi_b(\vk_1) \psi_c(\vk_2)  + \left[\omega \psi + \omega^2 +\pi  \right]_a  . 
\end{eqnarray}
Here index $a,b,c \in \{ 1, 2\}$, $i, j$ are spatial indices, and the Einstein summation notation is used. $\hat{\mL}_{ab} = \partial_{\eta} \delta_{ab} + \Omega_{ab} $ is the linear operator
with redefined time variable $d \eta = d\ln D(\tau)$, $D(\tau)$ is the linear growth rate,  $\Omega_{ab}$ (Eq. \ref{eqn:Omegaab})  is constant matrix at least in Einstein-de Sitter Universe. We also group all terms related to $\omega_i$ and $\pi_a = \{0, \pi_\theta=\nabla_i\pi_i \} $ in the bracket (Eq. \ref{eqn:fourier_dyn}) as they disappear in  the standard dust model.  With these terms included, it is clear that our equations are not closed. 
To simplify our expression later on, we will further rewrite above equations as 
\begin{eqnarray}
	\label{eqn:gen_dyn_sys}
	\partial_{\eta} \vpsi_{\vk}( \eta )  = \vchi_{\vk} \left( \vpsi, \vomega, \vpi;  \eta \right) . 
\end{eqnarray}
So here $\vchi_{\vk} =\{ \chi_{a} \}_{\vk} = \{ \chi_{1, \vk}, \chi_{2, \vk} \}$  \footnote{In the following, we will interchangeably use notation $\vchi(\vk)$ and $\vchi_{\vk}$ without further clarification.} 
includes the linear term $\Omega_{ab}\psi_b$, all nonlinear mode coupling terms $\gamma_{abc}\psi_b\psi_c$, and other terms related to $\omega_i$ and $\pi_a$ as well. 

Since we are mostly interested in the evolution of large-scale $\psi_a(\vk)$, an effective approach (e.g. the EFTofLSS) starts by separating these fields into large and small modes, with the help of some filter function $W_{\Lambda} (\vk)$, where $\Lambda$ is cutoff scale in $k$-space.  This function was conviniently chosen as the Gaussian smoothing in EFTofLSS. 
To avoid dealing with continuous field domain, here, we consider a finite cosmic volume $V$, with periodic boundary condition. Therefore, the number of total Fourier modes  $\delta(\vk)$ and $\vu(\vk)$ are countable\footnote{A full field formalism is certainly viable but beyond the scope of this paper. In that case, equation (\ref{eqn:PDF_tevol_text}) will be  integro-differential equation \citep{Blas15,Blas16}.  Here, we will make a simple assumption that our formalism is convertable to the field description by taking the limit $L \to \infty$ without questioning too much about the underlying mathematical structures. }.
In practice, however, we will not explicitly distinguish Fourier series and continuous transform in this paper.   For example, the Fourier space integration will {\it not} be replaced by series summation as  the difference would be negligible if $V$ becomes large enough.
Therefore, in this formalism, it is convenient to choose a sharp-k filter, and we could further select two seperate subsets of large-scale (soft) and small-scale (hard) modes. In the following, we will denote these two sets as 
\begin{eqnarray}
	\vpsi_{\Lambda} = \{ \vpsi(\vk) \}_{k<\Lambda} = \{  \delta(\vk), \theta(\vk) , \vomega(\vk)\}_{k<\Lambda},  \quad
	\vpsi_{\tildeLambda}  = \{ \vpsi(\vk) \}_{k>\Lambda} 
\end{eqnarray}
respectively. 

\begin{figure*}
	\includegraphics[width=1.\textwidth]{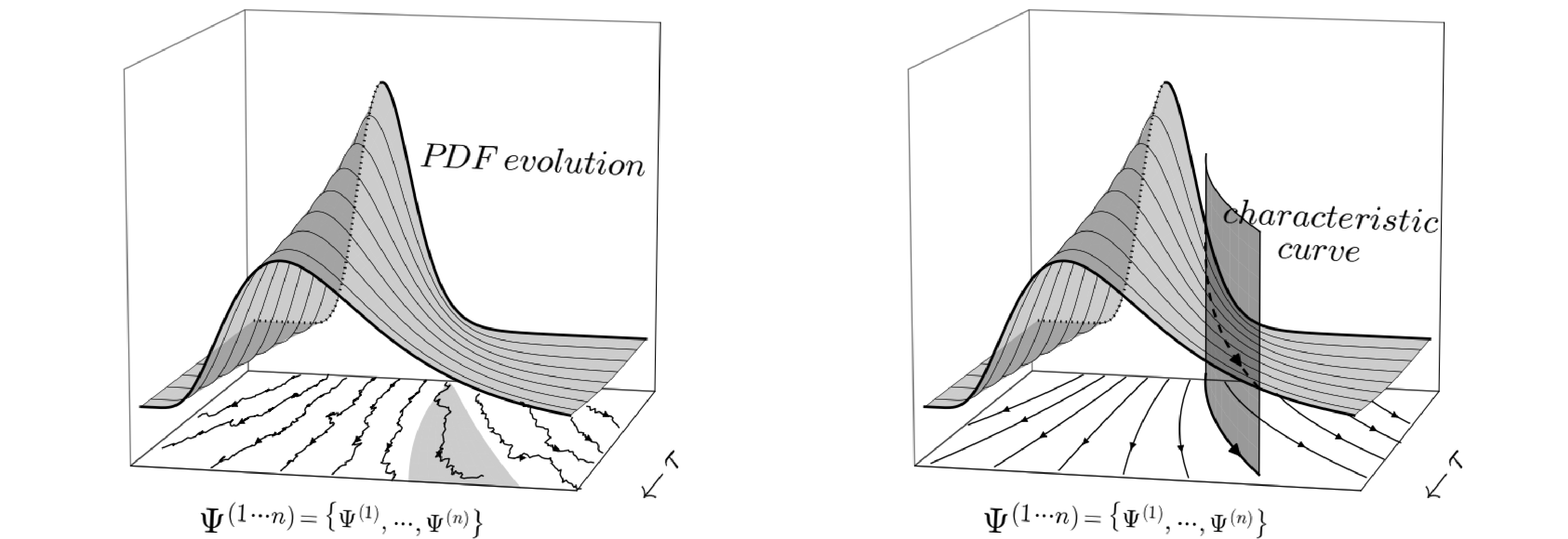}
	\caption{\label{fig:PDF_illustration} 
{\revision Illustration of the original dynamical system (equation \ref{eqn:gen_dyn_sys}) of soft modes $\vpsi_{\Lambda}=\{ \psi_{k^{(1)}} \cdots \psi_{k^{(n)}} \}$  on the {\it left}, and its effective solution (equation \ref{eqn:eff_dyn_general}) on the {\it right}. In each panel, the arrow indicates the temporal direction, and the grey two-dimensional surface represents the PDF evolution. 
By construction, the PDF evolution in both panel are identical, all originate from a Gaussian state and then evolve into some non-Gaussian form. On the floor of each figure, we also display individual trajectories of $\vpsi_{\Lambda}$ starting from given initial condition $\vpsi_{\Lambda, \ini}$. 
Since the original dynamics also depends on other random variables, i.e. $\vpsi_{\tildeLambda}$, $\vomega$ and $\vpi$, these trajectories appears somewhat `wiggly'. The grey band represents the probability transition function (\ref{eqn:tran_PDF_def}). 
On the right panel, the effective trajectories are obtained by projecting the `characteristic curves' (i.e. the dashed curve on the PDF surface) onto $\vpsi_{\Lambda}$ space. In general, these curves will behave differently.
 }
	}
\end{figure*}

{\revone
In the spirit of the effective theory, we are interested in the joint probability distribution function of the large-scale modes $\vpsi_{\Lambda}$:  }
$$\mP(\vPsi_{\Lambda}, \eta)  = \mP(\vpsi_{\Lambda} = \vPsi_{\Lambda}, \eta )= 
\mP(\{ \Delta_k, \Theta_k \}_{k<\Lambda}, \eta) . $$
Following the convention in turbulence literature \citep{Pope1985}, we will explicitly distinguish the statistical random variables (small letter) with the sample space variables (capitcalized, i.e. the one appears in the argument of probability function)\footnote{{\revision In probability theory, the sample space is the set of all possible outcomes of a random process. Therefore, unlike random variables, the parameters characterizing this space, i.e. the sample space variables, are not random \citep{probabilitybookANS}. } }. So here specifically,  $\vPsi, \Delta_k, \Theta_k $ are the sample space variables corresponding to the doublet field $\vpsi$, density contract $\delta_k$ and the divergence $ \theta_k$ respectively. 
Adopting the similar notation, we will also define the soft/hard modes parameter set as
$\vPsi_{\Lambda}= \{ \vPsi(\vk) \}_{k<\Lambda},  \vPsi_{\tildeLambda}= \{ \vPsi(\vk) \}_{k>\Lambda}$.

Needless to say that $\mP(\vPsi_{\Lambda}, \eta)$ is a super high-dimensional PDF, and there is little hope one will ever be able to describe its full shape. 
However, that does not mean we will not learn anything valuable. Following the PDF-based technique developed by \citep{Pope1985} in turbulence (see also its cosmology application in \citep{WS16}), one  could formally write down the time evolution equation of $\mP(\vPsi_{\Lambda}, \eta)$ as 
\begin{eqnarray}
	\label{eqn:PDF_tevol_text}
	\partial_{\eta} \mP (\vPsi_{\Lambda}; \eta)
	+  \sum_{a, ~ \vk} \partial_{\Psi_{a, \vk}}  \left[  \left \langle \chi_{a, \vk} 
	\middle |    \vPsi_{\Lambda} ; \eta \right \rangle   \mP (\vPsi_{\Lambda}; \eta)   \right] =0 , 
\end{eqnarray}
given the dynamical system (Eq. \ref{eqn:gen_dyn_sys}). 
Here $\left \langle \chi_{a, \vk} \middle |    \vPsi_{\Lambda} ; \eta \right \rangle $ is the conditional average given the values of $ \vPsi_{\Lambda} $. The second term sums over the partial derivatives with respect to $\Psi_{a, \vk}$ in our soft modes set $\vPsi_{\Lambda}$, so it is simply a high-dimensional divergence. 
{\revision As will be shown later, by writing down the above equation for $\mP (\vPsi_{\Lambda}; \eta)$, we have implicitly marginalized over small-scale modes.}
We further realizes that  Eq. (\ref{eqn:PDF_tevol_text}) is just the {\it conservation equation} of PDF $\mP(\vPsi_{\Lambda}, \eta)$ where the probability current ${\bm j}_{\mP} =  \left \langle \vchi \middle |    \vPsi_{\Lambda} ; \eta \right \rangle   \mP(\vPsi_{\Lambda}, \eta) $.

As a first order partial differential equation, Eq. (\ref{eqn:PDF_tevol_text}) could be reduced to a family of ordinary differential equations with the method of characteristics, which simply reads: 
\begin{eqnarray}
	\label{eqn:eff_dyn_general}
	\partial_{\eta} \vpsieff (\vk, \eta) = \left \langle  \vchi
	\middle |   \vpsieff; \eta  \right \rangle
\end{eqnarray}
To be schematically self-consistent, here we have already replaced the sample space variables $\vPsi_{\Lambda}$ with the dynamical variables $\vpsieff$. Compared to Eq. (\ref{eqn:gen_dyn_sys}), the conditional average term $ \left \langle  \vchi
\middle |   \vpsieff; \eta  \right \rangle$ will be different from $\vchi$ as long as we only consider soft modes.

The {\it key insight} of this method is to realize that, as the `solution' to the evolution equation of $\mP(\vPsi_{\Lambda}, \eta)$,  dynamical system Eq. (\ref{eqn:eff_dyn_general}) will produce exactly the same $\mP(\vPsi_{\Lambda}; \eta)$ at time $\eta$ with given initial condition $\mP(\vPsi_{\Lambda}, \eta=\eta_{\rm ini})$. 
This is more transparent when we try to re-derive the PDF equation (\ref{eqn:PDF_tevol_text}) from Eq. (\ref{eqn:eff_dyn_general}) instead of Eq. (\ref{eqn:gen_dyn_sys}).  From Eq. (\ref{eqn:eff_dyn_general}), it is easy to check that the derivation procedure in Sec. \ref{sec:eveq_PDF} would lead to the same PDF equation (\ref{eqn:PDF_tevol_text}) because 
$$ \left \langle \left \langle  \vchi	\middle |   \vpsieff  \right \rangle  \middle |  \vpsieff \right \rangle  =  \left \langle  \vchi	\middle |   \vpsieff  \right \rangle. $$
It is in this {\it statistical sense} that we call Eq. (\ref{eqn:eff_dyn_general}) the `{\it effective dynamics}' of the original system. 
{\revision  In Figure. (\ref{fig:PDF_illustration}), we schematically compare system (\ref{eqn:gen_dyn_sys}) and (\ref{eqn:eff_dyn_general}) on the left and right panel respectively.
In each panel, the arrow indicates the temporal direction, and the grey two-dimensional surface represents the PDF evolution. 
As just demonstrated, both systems share identical PDF history which originates from a Gaussian state and then evolve into some non-Gaussian form.
Individual trajectories are displayed on the floor of each panel. Since the original dynamics also depends on other random variables, i.e. $\vpsi_{\tildeLambda}$, $\vomega$ and $\vpi$, these trajectories appears somewhat `wiggly'.
On the contrary, the conditional average in equation (\ref{eqn:eff_dyn_general}) would lead to much smoother effective trajectories. As shown on the right panel, these effective trajectories are obtained by projecting the `characteristic curves' (dashed line on the PDF surface) onto $\vpsi_{\Lambda}$ space.
}

Now let us check the LSS dynamics in more details. Neglecting $\vpi$ and vorticity $\vomega$,  the dynamical system (\ref{eqn:fourier_dyn_full}) then returns to its standard form
\begin{eqnarray}
	\label{eqn:dust_model}
	\hat{\mL}_{ab} \psi_{b, \vk}  = \gamma_{abc}(\vk_1,\vk_2) \psi_{b,\vk_1} \psi_{c,\vk_2} .
\end{eqnarray}
Since the linear term  $\Omega_{ab} \psi_{b, \vk}$ will not be affected by the conditional average, the effective dynamics is expressed as 
\begin{eqnarray}
	\hat{\mL}_{ab} \psi_{b,\vk} 
	=  \gamma_{abc}(\vk_1,\vk_2) \left \langle \psi_{b,\vk_1} \psi_{c,\vk_2} \middle | 
	    \vpsieff; \eta \right \rangle ,  
\end{eqnarray}
where $|\vk|<\Lambda$, i.e. $\vpsi(\vk) \in \vpsieff$.

Since the large-scale modes $\vpsieff$ themselves could simply be taken out of the conditional 
average, the mode-coupling terms automatically separate into two groups: the soft-soft coupling  $$\langle \vpsieff \vpsieff | \vpsieff \rangle = \vpsieff \vpsieff $$
term, as well as the coupling involving hard modes, i.e. $ \left \langle \vpsi_{\tildeLambda} \vpsieff \middle | \vpsieff \right \rangle $ and $\left \langle \vpsi_{\tildeLambda} \vpsi_{\tildeLambda} \middle | \vpsieff \right \rangle $.  
So our effective dynamics simply reads
\begin{eqnarray}
	\label{eqn:eff_dyn_dm}
	\hat{\mL}_{ab} \psi_{b, \vk}  &=&  \bigl[ \gamma_{abc}(\vk_1,\vk_2) 
	\psi_{b,\vk_1} \psi_{c,\vk_2} \bigr]_{\Lambda\Lambda} 
	+  \mC_a(\vk, \vpsi_{\Lambda}; \eta), 
\end{eqnarray}
where $(\gamma\psi\psi)_{\Lambda\Lambda}$ denotes that the amplitude of wavenumbers 
$\vk_1$ and $\vk_2$ are both less than $\Lambda$. The extra contribution $\mC_a(\vk, \vpsieff; \eta)$ could be expressed as
\begin{eqnarray}
	\label{eqn:ct_def}
	\mC_a(\vk, \vpsi_{\Lambda}; \eta) &=& \int _{\tildeLambda\tildeLambda, 2\Lambda\tildeLambda} 
	d\vk_{12} ~\gamma_{abc}(\vk_1,\vk_2)   \left \langle   \psi_{b,\vk_1} \psi_{c,\vk_2} 
	\middle|  \vpsieff; \eta  \right \rangle . 
\end{eqnarray}
Here, the integration is taken over the Fourier region where at least one of $k_{1}, k_{2}$ is greater than $\Lambda$. It is easy to see that $\mC_a(\vk, \vpsi_{\Lambda}; \eta)$ plays a similar role as the effective stress tensor  $\tau_{ij}(\vx, \eta)$ introduced in EFTofLSS \citep{Baumann12,CHS12,H12,PZ13}. 
However, there is one important distinction, that is here $\mC_{a=0}(\vk, \vpsi_{\Lambda}; \eta)\ne 0$, so our formalism introduces an extra contribution to the continuity equation as well. Therefore, in real space, our effective equations read 
\begin{eqnarray} 
		\partial_{\tau} \delta_{\Lambda} + \nabla_i  \left [ (1+\delta_{\Lambda}) u_{\Lambda, i} \right] &=&  \widetilde{\mC}_0,  \nonumber \\
	\partial_{\tau} u_{\Lambda, i} + (u_{\Lambda, j} \nabla_j) u_{\Lambda, i}+ \mH u_{\Lambda, i} &=& 
	- {\nabla_i} \left( \frac{	\nabla^{-2}}{4\pi G \bar{\rho} a^2} \delta_{\Lambda}\right) + \nabla_i \nabla^{-2}  \widetilde{\mC}_{1}. 
\end{eqnarray}
This is because both conservation equations are nonlinear. 
{\revision EFTofLSS avoided such contribution by `linearizing' the continuity equation with the density-weighted  velocity $u^{(\rho)}_i = ( \sum_{\alpha} \rho_{\alpha} u_{\alpha, i} )/\bar{\rho}$ instead of the volume-weighted $u^{({\rm vol})}_i$ \citep{Mercolli_2014,Bertolini_2016}, here the summation $\sum_{\alpha}$ is taken over all particles at fixed Eulerian. }
Clearly from our previous derivation, this is also a viable choice if we prefer. However, as one goal of this paper is to understand our formalism and directly compare the standard perturbation theory, we will utilize the current volume-weighted velocity instead.

 \section{From the PDF Evolution to the Effective Dynamics} 
 \subsection{Evolution Equation of the Full Probability Function}
 \label{sec:eveq_PDF}


In this section, we would like to obtain Eq. (\ref{eqn:PDF_tevol_text}) from dynamical system (\ref{eqn:gen_dyn_sys}).  
{\revision The derivation follows the \cite{Klimontovich_1967} approach to kinetic theory, which was also adopted by \cite{Bertschinger_1995} in deriving the Vlasov equation. }
To derive the governing equation of the probability density function of this system,  we now consider an ensemble of the system.
For a single realization, the probability density function, i.e. the {\it fine-grained} PDF is described by the products of Dirac-$\delta$ functions  \footnote{We are only going to distinguish the notation of random variable and the sample space variable in this section as it clarify the derivation. However, we will not make such distinction anywhere else in the paper.}
\begin{eqnarray}
\label{eqn:finegrain_P}
\mP_f \left (\vPsi_{\Lambda}; \eta \right ) = 
\prod_{\vk} \delta_D \left [\vPsi_{\vk} - \vpsi_{\vk}(\eta)  \right ] 
= \prod_{\alpha, \vk} \delta_D \left [\Psi_{\alpha, \vk} - \psi_{\alpha, \vk}(\eta)  \right ], 
\qquad \vPsi_{\vk} \in \vPsi_{\Lambda}, 
\end{eqnarray}
where  $\vPsi_{\vk}$ is the sample space variable corresponds to $\vpsi_{\vk}$. 
Now let us consider an ensemble of  systems (\ref{eqn:gen_dyn_sys}). 
There are certain freedom in choosing the ensemble, for example in \cite{WS16}, it is a 
sample of Lagrangian fluid elements, or equivalently the density-weighted field in Eulerian space, 
but here we assume it is different realizations of our Universe.
By definition, the PDF of the ensemble could be obtained by taking average of the fined-grained 
PDF \citep{Pope1985},  i.e.
\begin{eqnarray}
\label{eqn:PDF_def}
\left \langle \mP_f  \left (\vPsi_{\Lambda} ; \eta \right )  \right \rangle  =  
\int d\vPsi^{\pri} \mP(\vPsi^{\pri}; \eta) \delta_D(\vPsi^{\pri}-\vPsi_{\Lambda})
=\mP \left (\vPsi_{\Lambda} ; \eta \right )  .
\end{eqnarray}
To proceed, one takes the time derivative of $\mP \left (\vPsi_{\Lambda}; \eta \right) $,  
\begin{eqnarray}
\label{eqn:Pn_evol_general}
\partial_{\eta} \mP  \left (\vPsi_{\Lambda}; \eta \right )  
&=& \left \langle \partial_{\eta}  \mP_f \left (\vPsi_{\Lambda}; \eta \right )   \right \rangle  
 =   \sum_{a, ~ \vk} \left \langle  \left[  \partial_{\eta} \psi_{a, \vk} (\eta) \right]   \partial_{\psi_{a, \vk}} \mP_f  
 \left(\vPsi_{\Lambda}; \eta \right )    \right \rangle  \nonumber \\ 
    &=& -  \sum_{a, ~ \vk} \left \langle  \left[  \partial_{\eta} \psi_{a, \vk} (\eta) \right]  
     \partial_{\Psi_{a, \vk}} \mP_f    \right \rangle  
    = -  \sum_{a, ~ \vk} \partial_{\Psi_{a, \vk}}  \left[  \left \langle \chi_a 
    \left (\vpsi_{\vk}, \vomega, \vpi; \eta \right ) 
    \mP_f  \right \rangle     \right] . 
\end{eqnarray}
In the second last equality, we have used definition (\ref{eqn:finegrain_P}) and the fact that 
$\partial_{\psi_{a, \vk} } \mP_f = - \partial_{\Psi_{a, \vk} }  \mP_f $. And since the sample space variables $\Psi$ are not random, we can take the derivative out of the average.  To further estimate $\langle \vchi \mP_f \rangle $, one notices that $\vchi$ is also random as it is just a function of $\{  \vpsi_{\Lambda}, \vpsi_{\tilde{\Lambda}}, \vpi, \vomega \} $ etc.
Assuming $\{ \vchi, \vpsi \}$ follows the joint-PDF $\mP(\vX, \vPsi; \eta)$, we have
\begin{eqnarray}
	\left \langle  \chi_{a} \mP_f \right \rangle &=&  \int d\vPsi^{\pri} d\vX^{\pri} X_{a} \delta_D(\vPsi^{\pri}-\vPsi_{\Lambda}) 
		 \mP (\vX^{\pri}, \vPsi^{\pri}; \eta) \nonumber\\
		 &=& \int d\vX^{\pri} X_{a} \mP(\vX^{\pri}| \vPsi_{\Lambda}; \eta ) \mP(\vPsi_{\Lambda}; \eta)
		  = \left \langle   \chi_{a} \middle | \vPsi_{\Lambda}; \eta \right \rangle \mP(\vPsi_{\Lambda}; \eta).
\end{eqnarray}
Here $X_{a}$ is the sample space variable of $\chi_{a}$.  And eventually one obtains  the continuity equation of $\mP(\vPsi; \eta)$ \cite{Blas15,Blas16}
\begin{eqnarray}
\label{eqn:PDF_tevol_formal_deriv}
\partial_{\eta} \mP \left (\vPsi_{\Lambda}; \eta \right )  +  \sum_{a, ~\vk} \partial_{\Psi_{a, \vk}}  
\left[  \left \langle \chi_{a} 
 \middle |    \vPsi_{\Lambda}; \eta \right \rangle   \mP \left (\vPsi_{\Lambda}; \eta \right )   \right] =0 .
\end{eqnarray}
So as long as one could know {\it in advance} the averaged $\vchi$ given the constraints of $\vPsi_{\Lambda}$, this is a closed equation of $\mP(\vPsi_{\Lambda}; \eta)$. 

As the first order partial differential equation, one could apply the so-called method of characteristic to reduce the problem into a family of ordinary differential equations:
\begin{eqnarray}
	\label{eqn:chartraj}
\partial_{\eta} \vPsi (\vk, \eta) = \left \langle  \vchi
\middle |   \vPsi_{\Lambda}; \eta  \right \rangle , \qquad |\vk | < \Lambda. 
\end{eqnarray}
Unlike  in Eq. (\ref{eqn:eff_dyn_general}), here we expressed in the sample space variable to emphasize their relation to Eq. (\ref{eqn:PDF_tevol_formal_deriv}). 
As explained in Figure. (\ref{fig:PDF_illustration}), these are the trajectories along which the probability $\mP(\vPsi_{\Lambda}; \eta)$ will be conserved. 
The physical meaning of these trajectories becomes much clearer when we further define the 
{\it probability current} along the direction of $ \vPsi_{\Lambda}$  in equation (\ref{eqn:PDF_tevol_formal_deriv})
\begin{eqnarray}
	\label{eqn:PDFcurrent}
	{\bm j}_{\mP}  (\vPsi_{\Lambda}; \eta) &=&  \vu_{\mP} \mP (\vPsi_{\Lambda}; \eta) =  
	\left(  \partial_{\eta} \vPsi_{\Lambda}  \right) \mP  (\vPsi_{\Lambda}; \eta) 
	= \left \langle \vchi \middle | \vPsi_{\Lambda} ; \eta \right \rangle  \mP  (\vPsi_{\Lambda}; \eta).   
\end{eqnarray}
where $\vu_{\mP} = \partial_{\eta} \vPsi_{\Lambda}$ could be understood as the `velocity' in the parameter space.

To better understand the difference between characteristic trajectories and the original dynamics, it is helpful to consider the probability transition function, which is defined as the conditional  probability of $\vpsi_{\Lambda}$  for a given initial state
$\vpsi_{\Lambda, \ini}$, i.e. 
\begin{eqnarray}
	\label{eqn:tran_PDF_def}
	\mT  \left[ \vPsi_{\Lambda} \middle |  \vPsi_{\Lambda, \ini}  \right] =
	\frac{ \mP\left[  \vPsi_{\Lambda}, \vPsi_{\Lambda, \ini} \right] }
	{\mP \left[  \vPsi_{\Lambda, \ini} \right ] }. 
\end{eqnarray}
The PDF at epoch $\eta$ could then be expressed as an integral over all possible initial conditions
\begin{eqnarray}
	\mP \left[ \vPsi_{\Lambda}; \eta \right] &=& \int \left( \mD \vPsi_{\Lambda, \ini}  \right)
	~ \mP\left [ \vPsi_{\Lambda, \ini}  \right] 
 \mT \left[ \vPsi_{\Lambda}  \middle |  \vPsi_{\Lambda, \ini}   \right]  .
\end{eqnarray}
So for a deterministic system whose dynamics is fully described by $\vpsi_{\Lambda}$, the current is simply the original dynamical equations (\ref{eqn:gen_dyn_sys}), and the probability transition function $\mT$ is just a Dirac $\delta$ function i.e. $\delta_D\left (\vPsi_{\Lambda}-\vPsi_{\Lambda, \ini} \right )$.
However, in many cases, $\chi^{(i)}$ would also depend on random variables other than $\vpsi_{\Lambda}$, as here it also depends on $\{ \vpsi_{\tilde{\Lambda}}, \vomega, \vpi\}$.  Therefore, the transition probability $\mT$ will generally be broadened, illustrated as shaded region on the floor of the left panel in Figure.~ (\ref{fig:PDF_illustration}). 
In such situation, the effective trajectory Eq. (\ref{eqn:chartraj}) describes the averaged flow of the PDF for given initial $\vPsi_{\Lambda, 0}$. 
Finally, to be consistent with our notation, we then rewrite equation (\ref{eqn:chartraj}) using dynamical variables  $\vpsieff$, so that we eventually obtain equation (\ref{eqn:eff_dyn_general}). 


\subsection{Effective Dynamics of LSS}
\label{sec:setdustmodel}
Following the earlier derivation, we could apply the effective dynamical equation (\ref{eqn:eff_dyn_general}) to the large-scale structure. 
In this section, we will first focus on the dust model, and then generalize to the orbit-crossing in the next section.  
In both cases, we are interested in  the statistical evolution of the density contrast $\delta$ and  peculiar velocity $\vu$ field, which is encoded in the joint PDF of $\delta$ and $\vu$ . 
Substituting cosmic dynamics to equation (\ref{eqn:eff_dyn_general}), we have the effective dynamics of LSS as
\begin{eqnarray}
	\hat{\mL}_{ab} \psi_{b,\vk} 
	=  \gamma_{abc}(\vk_1,\vk_2) \left \langle \psi_{b,\vk_1} \psi_{c,\vk_2} \middle | 
	\vpsieff; \eta \right \rangle .
\end{eqnarray}
Here we are integrating over $\vk_1, \vk_2$, since the large scale modes $ \vpsi_{\vk} \in \vpsieff$ could simply be taken out of the conditional  average, we automatically separates the quadratic mode coupling terms into two groups
\begin{eqnarray}
	\label{eqn:eff_dyn_dm}
	\hat{\mL}_{ab} \psi_{b, \vk}(\eta)  &=&  \bigl[ \gamma_{abc}(\vk_1,\vk_2) 
	\psi_{b,\vk_1}(\eta) \psi_{c,\vk_2} (\eta)\bigr]_{\Lambda\Lambda} 
	+  \mC_a(\vk, \vpsi_{\Lambda}; \eta), 
\end{eqnarray}
where $(\gamma\psi\psi)_{\Lambda\Lambda}$ denotes that the amplitude of wavenumbers 
$\vk_1$ and $\vk_2$ are both less than $\Lambda$. The extra contribution $\mC_a(\vk, \vpsieff; \eta)$ could be expressed as
\begin{eqnarray}
	\label{eqn:ct_def}
	\mC_a(\vk, \vpsi_{\Lambda}; \eta) &=& \int _{\tildeLambda\tildeLambda, 2\Lambda\tildeLambda} 
	d\vk_{12} ~\gamma_{abc}(\vk_1,\vk_2)   \left \langle   \psi_{b,\vk_1} \psi_{c,\vk_2} 
	\middle|  \vpsieff; \eta  \right \rangle . 
\end{eqnarray}
Here, the integration is taken over the Fourier region where at least one of $k_{1}, k_{2}$ is greater than $\Lambda$. 
{\revision As will be discussed later in Sec. \ref{sec:effterm_measure}, this equation might suggest an interesting way to directly measure these effective terms from simulation. }
Similar to equation (\ref{eqn:form_solution_dm}), the formal solution of $\psi_a(\vk, \eta)$ could then be expressed as the linear propagation of source terms at the right hand side of equation (\ref{eqn:eff_dyn_dm})
\begin{eqnarray}
	\label{eqn:nonlinear_solu_C}
\psi_{a,\vk}(\eta) &=& g_{ab}(\eta) \phi_b(\vk) + \int_0^{\eta} ds ~g_{ab}(\eta-s)  \bigl[ \gamma_{bcd}(\vk_1,\vk_2) 
  \psi_{c,\vk_1}(s)   \psi_{d,\vk_2}(s) \bigr]_{\Lambda\Lambda} \nonumber \\
&& +  ~\int_0^{\eta} ds ~g_{ab}(\eta-s)   \mC_b(\vk, \vpsieff; s) . 
\end{eqnarray}
To simply the expression,  we will also denote the last term  as
\begin{eqnarray}
	\label{eqn:defmS}
 \mS_a(\vk, \eta) = \int_0^{\eta} ds~ g_{ab}(\eta-s) \mC_b(\vk, \vpsieff; s). 
\end{eqnarray}

From the effective theory point of view, as long as one could accurately fit or estimate the value 
of $\mC_a(\vk, \vpsieff; \eta)$, the entire statistical information, including all orders of polyspectra of $\vpsi_{\Lambda}$ would be recovered precisely, given that the large-scale mode coupling $[\gamma \psi\psi]_{\Lambda\Lambda}$ is much easier to calculate. 
{\revision Now we would like to better understand the coefficients $\mC_a(\vk, \vpsieff; \eta)$ so that some form of parametrization or expansion could be adopted.
From the definition Eq. (\ref{eqn:ct_def}), it is clear that $\mC_a(\vk, \vpsieff; \eta)$ must be a function of  {\it all} soft modes $\vpsieff$ at time $\eta$. Therefore, we can at least Taylor expand it in $\vpsieff(\eta)$.  
A more rigorous treatment is to write down the conditional average in Eq. (\ref{eqn:ct_def}) as the integration over the joint probability density function of all relevant variables $\vGamma=\{ \vpsi_{\Lambda}, \vpsi_{\tildeLambda} \}$. By definition, it is expressed as }
\begin{eqnarray}
\label{eqn:condav_def}
\langle \psi_{b,\vk_1} \psi_{c, \vk_2} | \vpsi_{\Lambda} \rangle \mP(\vpsi_{\Lambda}) =
 \int \mD \vpsi_{\tildeLambda}  ~ (\psi_{b,\vk_1} \psi_{c, \vk_2}) \mP(\vGamma), 
\end{eqnarray}
where $\mD \vpsi_{\tildeLambda} $ is the volume element of the $\vpsi_{\tildeLambda}$-space.
Next, we can apply the Gram-Charlier expansion, and substitutes the non-Gaussian PDF $\mP(\vGamma)$ in Eq. (\ref{eqn:condav_def})  with a series of Gaussian PDF $\mP_G$ and its derivatives \cite{C67,JWACB95,Amen96,BM98}

\begin{eqnarray}
	\label{eqn:GC_expansion}
	\mP(\vGamma) = \mP_G(\vGamma) \left [ 1+ \sum_{n\ge 3} \frac{1}{n!} \langle \vGamma^n \rangle_{GC}
	\otimes_T  {\bm H}_n (\vGamma)   \right]. 
\end{eqnarray}
Here $\otimes_T$ denotes the inner produce of two tensors, $\mP_G$ is the Gaussian part of $\mP$, 
and the Hermite tensor ${\bm H}_n$ is 
defined as ${\bm H}_n(\vGamma) = (-1)^n \mP_G^{-1}(\vGamma) \partial^n \mP_G(\vGamma) /\partial^n \vGamma$. 
The conditional average could then be expressed in terms of Gram-Charlier coefficients 
$\langle \vGamma^n \rangle_{GC}$ which then related to the cumulants of various modes, 
i.e. the polyspectra here.  
Therefore, we have an expansion of $\mC_a$ as
\begin{eqnarray}
\label{eqn:ct_taylor_exp}
 \mC_a(\vk,\vpsieff) &=& \frac{\partial  \mC_{a,\vk}}{\partial \psi_{b, \vk} } \psi_{b, \vk} + 
 \frac{1}{2!}\frac{\partial^2  \mC_{a,\vk}}{\partial \psi_{b,\vk_1} \partial \psi_{c,\vk_2}}\psi_{b,\vk_1}  
 \psi_{c,\vk_2}  
  +  \frac{1}{3!}\frac{\partial^3  
 \mC_{a,\vk}}{\partial \psi_{b,\vk_1} \partial \psi_{c,\vk_2} \partial \psi_{d,\vk_3}}
   \psi_{b,\vk_1} \psi_{c,\vk_2} \psi_{d,\vk_3}   \nonumber \\
   &&  + ~ \cdots
\end{eqnarray}
where $\vk, \vk_1, \cdots, \vk_n \in \vpsi_{\Lambda}$, and we have assumed $\langle \psi_{a, \vk} \rangle = 0$.
{\revision
Eq. (\ref{eqn:ct_taylor_exp}) suggests a very different expansion scheme compared with the EFTofLSS, where the stress tensor is expanded in functions that are spatially local but temporally non-local.
On the contrary, as the equal time PDF and $\mC^{(m)}(\vk, \vpsieff; \eta)$ are functions of $\vpsieff(\eta)$, our natural expansion basis is {\it non-local in Fourier space} but {\it temporally local}.  
As demonstrated in Figure. (\ref{tab:SPT_comp}) and Sec. \ref{sec:PT}, the Fourier space non-locality here is essential to recover the SPT at one-loop order. 
Of course, it is always possible to further re-expand these terms the same way EFTofLSS does, and it would be interesting to examine the consequences. We will defer this for future studies.
}


\subsection{The Orbit Crossing and Stochasticity}
\label{sec:orbit_crossing}
In the standard theory of the structure formation, before the orbit-crossing, the matter is considered to occupy a three-dimensional sheet in the six-dimensional phase space.  Consequently, both vorticity $\omega_i$ and stress tensor $\sigma_{ij}$ vanish until the orbit crossing when this sheet starts to wind up. After the orbit crossing, however, both quantities start to emerge \cite{PB99,WS14}.
As shown in equation (\ref{eqn:cdyn_eul}), the challenge is that the system is not closed. One simple solution is to supplement their information from simulation measurement.
For example, \cite{PS09} demonstrated that at the leading order, the dominant effect comes from $\pi_{\theta}$.
They  showed that the correction to the power spectra at the leading order would be \cite{PS09}
$P_{aa}(\vk,\eta) =  P_{dust, aa}(\vk,\eta) +  P_{a \pi_{\theta}} (\vk,\eta) / \left [(n/2-1) ( n/2+3/2) \right] $, 
where $n$ is defined as the growth index of $\pi_{\theta}$ so that  $ \pi_{\theta} \propto D^{n/2}$. From the direct measurement of the simulation, this would cause $1\%$ correction to $P_{\theta\theta} (P_{\delta\delta})$ around $k \approx 0.1 (0.2)~{\rm Mpc/h}$.

From Eq. (\ref{eqn:eff_dyn_general}) and its derivation, the effective trajectories do not distinguish between small-scale perturbation and other variables, and the extra terms $\pi_a(\vk, \eta)=\left \{0,   \pi_{\theta}(\vk)  \right \}$ and $\omega_i$ can be substituted with corresponding conditional averages as well. Thus the effective system now reads 
\begin{eqnarray}
	\label{eqn:eff_traj_oc}
	\hat{\mL}_{ab} \psi_{b,\vk} 
	& =&  \gamma_{abc}(\vk_1,\vk_2) \left \langle \psi_{b,\vk_1} \psi_{c,\vk_2} \middle | \vpsi_{\Lambda} \right \rangle 
	+  \gamma^{\omega}_{aib}(\vk_1,\vk_2) 
	\left \langle \omega_{i,\vk_1} \psi_{b,\vk_2} \middle | \vpsi_{\Lambda} \right \rangle  \nonumber \\
	&& \times  + \gamma^{\omega^2}_{aij} (\vk_1, \vk_2)  
	\left \langle  \omega_{i, \vk_1} \omega_{j, \vk_2} \middle |  \vpsi_{\Lambda} \right \rangle  
	+  \left \langle \pi_{a, \vk} \middle |  \vpsi_{\Lambda} \right \rangle  . 
\end{eqnarray}
All these extra terms could to be directly calibrated from simulation, and we will postpond such measurement for future study.

\section{The Perturbation Theory}
\label{sec:PT}
After introducing the framework, we would like to understand the small-scale effective term $\mC_a(\vk, \vpsi_{\Lambda}; \eta)$	in more details. Practically, this could be measured from the simulation.  However, one could also gain some insight from analytical study as well. 
Since by construction, the effective dynamics produces exactly the same statistics as the true LSS dynamics.  Thus, if we choose to perturbatively expand those terms $\mC_a$, our formalism should produce the same result as SPT.  But this is not obvious at first glance since $\mC_a$ is expressed as the conditional average. 
In this section, we will further discuss various ways one could expand these contributions and the consequence in constructing the effective dynamics.

\begin{table}[t]
	\begin{center}
		\begin{tabular}{|c|c|}
			\hline
			$P^{SPT}_{1-loop}(k)$ &       Effective Terms ({\it ET})  \nonumber \\
			\hline
			$P_{13, \Lambda\Lambda}(k), P_{22, \Lambda\Lambda}(k)$ & 
			{\it no ET}  \nonumber\\
			\hline
			$P_{13, \tildeLambda\tildeLambda}(k)$ & $\mC^{B^T (\Rmnum{1})}_{\tildeLambda\tildeLambda, a}(\vk)$ ,
			$\mC^{B^T (\Rmnum{2})}_{\tildeLambda\tildeLambda, a}(\vk)$  \nonumber\\
			$P_{22, \tildeLambda\tildeLambda}(k)$ & $\mC^{B^T (\Rmnum{3})}_{\tildeLambda\tildeLambda, a}(\vk)$  \nonumber\\
			\hline
			\quad $P_{13, \Lambda\tildeLambda}(k)$ {\rm hard-integral}  \quad & 
			\quad $\mC^{B^T (\Rmnum{1})}_{2\Lambda\tildeLambda, a}(\vk,  \vk-\vq, \vq-\vk)$  \quad\nonumber \\
			\quad $P_{13, \Lambda\tildeLambda}(k)$ {\rm soft-integral} \quad & 
			\quad $\mC^{B^T (\Rmnum{2})}_{2\Lambda\tildeLambda, a}(\vk,  \vk-\vq, \vq-\vk)$ \quad\nonumber \\
			\quad$P_{22, \Lambda\tildeLambda}(k)$ \quad & 
			\quad $\mC^{B^T (\Rmnum{3})}_{2\Lambda\tildeLambda, a}(\vk,  \vk-\vq, \vq-\vk)$ \quad \nonumber \\
			\hline
		\end{tabular}
	\end{center}
	\caption{\label{tab:SPT_comp}
 One-loop expansion of $\mC(\vk)$ recovers the result of standard perturbation theory. One could find the detailed calculation in Appendix \ref{sec:1lp_PT}. We show every SPT 1-loop terms and their corresponding $\mC(\vk)$. Here the superscripts $B^{T}(\Rmnum{1})$,   $B^{T}(\Rmnum{2})$ and $B^{T}(\Rmnum{3})$ denote three different contributions of the tree-level bispectra (Eq. \ref{eqn:Btree_sum}).  As discussed in main text, the perturbative expansion of these effective terms are just a re-organization of SPT diagrams, and it is essential to include all these non-local terms for the complete recovery of SPT. 
	}
\end{table}

\subsection{Effective Coefficients at One-loop Order}

To better understand these coefficients, we explicitly calculate them at one-loop order and present all technical details in Appendix \ref{sec:1lp_PT}. As already mentioned, we could apply the so-called Gram-Charlier series to expand the non-Gaussian PDF $\mP(\vGamma)$ in terms of Gaussian $\mP_G(\vGamma)$ and calculate the conditional average $ \left \langle \vpsi_{\vk_1} \vpsi_{\vk_2} \middle |  \vpsi_{\Lambda} \right \rangle $ in Eq. (\ref{eqn:ct_def}).
As shown in Appendix \ref{app:condexp}, this term appears differently depending on the integrating range of $\vk_1$ and $\vk_2$.  
If both $k_1$ and $k_2$ are greater than the cut-off scale $\Lambda$, i.e. considering the {\it hard-hard} coupling,  $\mC^{B}_{\tildeLambda\tildeLambda,a}(\vk, \eta)$  is then expressed as an integral of the bispectrum with the kernel $\gamma_{abc}$
\begin{eqnarray}
	\label{eqn:stpara_B_maintext}
	\mC^{B}_{\tildeLambda\tildeLambda,a}(\vk; \eta) &=& \int_{\tildeLambda\tildeLambda} 
	d\vq ~ \gamma_{abc} (\vq, \vk-\vq)  B_{bcd} (\vq, \vk-\vq, -\vk; \eta)  
	P^{-1}_{de}(k, \eta)  \psi_e(\vk, \eta). 
\end{eqnarray}
Here $\tildeLambda\tildeLambda$ denotes the integration region  $q>\Lambda$ and  $|\vk-\vq|>\Lambda$. All the quantities here are evaluated at time $\eta$.  The bispectrum $B(\vq, \vk-\vq, -\vk)$  should be fully non-linear, as well as the power spectra $P_{de}(k, s)$ and its inverse.  

Following the integrated nonlinear solution (Eq. \ref{eqn:nonlinear_solu_C} and \ref{eqn:defmS}) \citep{BCGS12review,CS06a,CS06b},  the time-evolved non-linear contribution of $\mC$ is
\begin{eqnarray}
	\mS^B_{\tildeLambda\tildeLambda, a}(\vk; \eta) = \int_0^{\eta} ds ~ g_{ab}(\eta -s )  
 \int_{\tildeLambda\tildeLambda}  d\vq ~ \gamma_{abc} (\vq, \vk-\vq)  B_{bcd} (\vq, \vk-\vq, -\vk; s)  
P^{-1}_{de}(k, s)  \psi_e(\vk, s) . \nonumber \\
\end{eqnarray}
To the lowest order, one could substitute the non-linear $B_{abc}$ with three different contributions of tree-level bispectra, which we will denote as $B^{T(I)}$, $B^{T(II)}$ and $B^{T(III)}$ respectively (see Eq. \ref{eqn:Btree_sum}-\ref{eqn:Btree_cyc}, diagrams are shown in Figure. \ref{fig:tree_bispectrum}). 
Following the diagram convention introduced by \citep{CS06a},  we present the time evolved $\mS_a(\vk, \eta)$ in Figure. (\ref{fig:P1l_bi}). The only new component we introduce here is the inverse power spectrum $ \otimes^{-1} =P^{-1}_{ab} (\vk)$, which obeys the relation $ \otimes^{-1}  \otimes = P_{ac} (\vk) P^{-1}_{cb}(\vk) = \delta^K_{ab} $.  
As shown, our diagram simply connect the kernel $\gamma_{abc}$ with bispectrum $B^{T}$ and $P^{-1}$.   Since the kernel $\gamma_{abc}(\vq, \vk-\vq)$ is symmetric with respect to $\vq$ and $\vk-\vq$,  the contribution from $B^{T (\Rmnum{1})}$ and $B^{T (\Rmnum{2})}$ are identical.  Therefore, there are only two distinct contributions to $\mS_a$, which are shown in Figure. (\ref{fig:P1l_bi}).
For people who are familiar with SPT, these two diagrams are identical to $P^{(13)}$ and $P^{(22)}$ in SPT respectively. 
Similarly the effective term of {\it soft-hard} ($\Lambda\tildeLambda ~and ~ \tildeLambda\Lambda$) coupling could be expressed as
\begin{eqnarray}
	\mC^{B}_{2\Lambda\tildeLambda,a}(\vk, \eta) &=&
	2 \int_{\Lambda\tildeLambda} d\vq ~ \gamma_{abc}(\vq, \vk-\vq)
	B_{bcd}(\vq,  \vk-\vq, -\vk; \eta)   P^{-1}_{ce}(|\vk-\vq|; \eta)  P^{-1}_{df}(k; \eta)  \nonumber \\
	&& \times \psi_{e}(\vq-\vk; \eta) \psi_f(\vk; \eta)   \psi_g(\vk-\vq; \eta). 
\end{eqnarray}
As shown in Appendix \ref{sec:PT_SHcoupling}, when considering the power spectrum, this will also produce the same SPT $P^{(13)}$ and $P^{(22)}$ in {\it soft-hard} coupling region.  In Table (\ref{tab:SPT_comp}), we summarize all contributions of $\mS_a$ and its relation to one-loop SPT. 
It is essential to note that we were only able to recover the SPT result by including all non-local terms.

\subsection{General Expansion of Effective Coefficients}
\label{sec:Cexpansion}
\begin{figure*}
	\includegraphics[width=\textwidth]{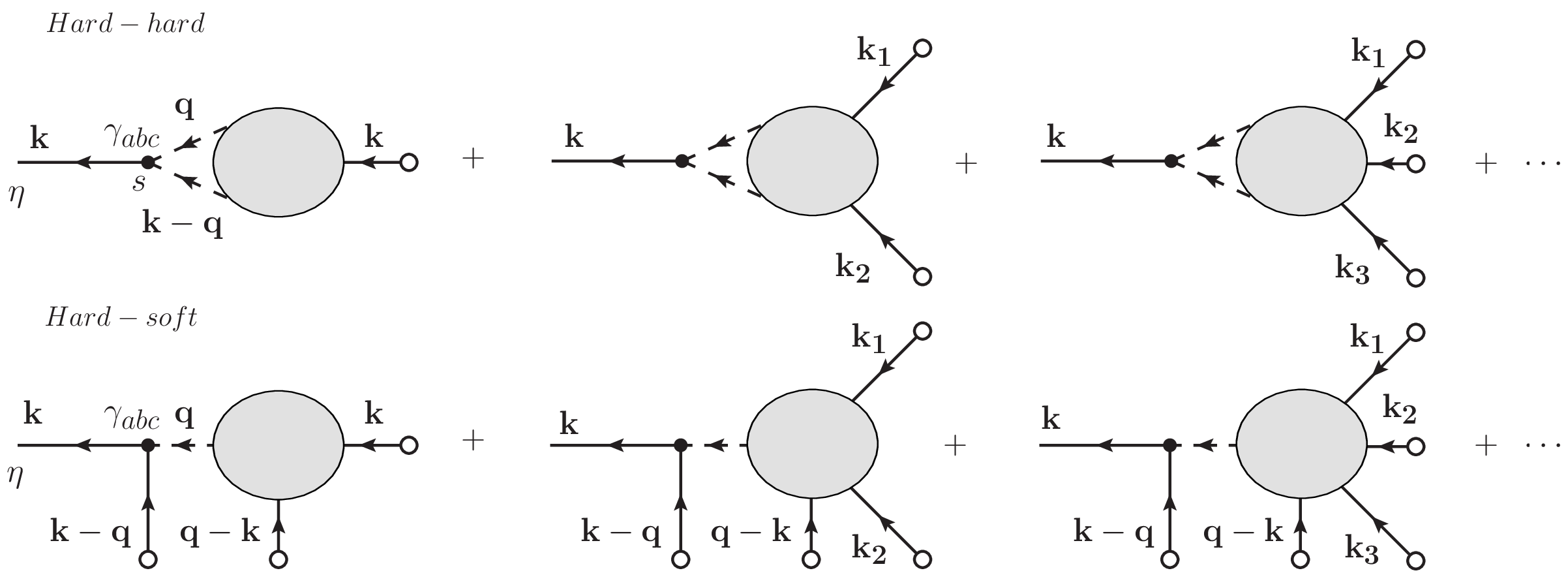}
	\caption{\label{fig:eff_coefficients_general}
		Diagrams for effective coefficients $\mS_a(\vk, \Lambda)$, which could be expressed as integral over 
		higher-order correlators, e.g. bispectrum, trispectrum etc., with the kernel $\gamma_{abc}$. 
		Here, the solid line represents the linear growth of large-scale mode $\vpsi_{\Lambda}$, 
		while the dashed line corresponds to small-scale modes $\vpsi_{\tildeLambda}$.  
		The grey ellipse symbolize complicated non-linear interactions among various modes. 
		Ideally, these interactions could be measured and carefully calibrated with simulation, which then leads to an effective theory of large-scale structure.
		In the first row, we illustrate the hard-hard coupling and show the hard-soft interaction at the second row.
	}
\end{figure*}

As discussed in Sec. {\ref{sec:setdustmodel}},  we can in general expand  $\mC_a(\vk, \eta)$  in terms of $\psi_a(\vk, \eta)$, i.e.
\begin{eqnarray}
\label{eqn:Cexpansion_short}
 \mC_a(\vk, \eta)=  \sum_n  \mC^{(n)}_a(\vk,  \eta) = \sum 
 \partial^{n}\mC_{a b_1 \cdots b_n } \left [  \vpsi_{b_1, k_1}  \cdots \vpsi_{b_n, k_n} \right].
\end{eqnarray}
Here the final summation is over all orders and possible combinations of $\vk_1 \cdots \vk_n$ as well. We have also adopted  the short-handed notation where the n-th order derivative ${\partial^{n} \mC}$ denotes
\begin{eqnarray}
	\label{eqn:dCm}
	\partial^{n} \mC_{a b_1\cdots b_n} =  \frac{1}{n!} \frac{\partial^n \mC_a}{\partial \psi_{b_1, k_1} \cdots 
		\partial \psi_{b_n, k_n}    }  , 
\end{eqnarray}
and they are some non-linear functions of all  large-scale modes $\vpsi_{\Lambda}$.
With this type of expansion, the effective terms are therefore intrinsically non-local in Fourier space. In the discrete limit where the number of large-scale modes are finite,  one eventually obtains a set of coupled differential equations  
\begin{eqnarray}
	\label{eqn:eff_eq_motion}
	\hat{ { \mL}}  \vpsi_{\vk} &=&   [ \vgamma \vpsi_{\Lambda} \vpsi_{\Lambda} ]_{\vk} + 
	\sum \left [ {\partial^n \mC} \otimes \left ( \vpsi_{\Lambda} \cdots \vpsi_{\Lambda} \right) \right ]_{\vk} 
\end{eqnarray}
where $\otimes$ is inner product, and $[ \cdots ]_{\vk}$ denotes all possible couplings where $\sum_i \vk_i = \vk$. 
In Figure.~ (\ref{fig:eff_coefficients_general}), we illustrate the first several non-perturbative diagrams 
representing the time-evolved effective terms $\mS_a(\vk, \vpsi_{\Lambda}; \eta)$. 
Notice that two diagrams in the first column are nonlinear generalization of the one-loop effective
terms (see Figure. \ref{fig:P1l_bi} and \ref{fig:hard_soft_1lp}). 
For example, the first hard-hard diagram consists of  equation (\ref{eqn:stpara_B}) with {\it non-linear} bispectrum $B_{\rm nl}(\vq,\vk-\vq,-\vk)$, but it also includes similar integrals with even more points polyspectra, e.g. the second and third column in Figure. (\ref{fig:eff_coefficients_general}).

For the hard-hard coupling ({\it the first row} in Figure. \ref{fig:eff_coefficients_general}), the linear effective term (with regard to the number of soft modes)  in general depends on $\vk$, 
\begin{eqnarray}
\label{eqn:linear_effterm}
\mC^{(1)}_{\tildeLambda\tildeLambda,a}(\vk)  =  \partial^{1} \mC_{\tildeLambda\tildeLambda, ab} 
(\vk) \psi_b(\vk) = c_s^2(\vk) k^2  \psi_a(\vk),
\end{eqnarray}
where $\partial^1 \mC$ is defined in equation (\ref{eqn:dCm}). 
To be consistent with EFTofLSS,  we also name the coefficient as the ``sound speed'' $c_s^2(\vk)$ in the last equality. As one would expect, $c_s^2(\vk)$ is generally scale-dependent. 
This corresponds to the first diagram, all other diagrams depend explicitly on soft modes other than $\vk$.
For the soft-hard couplings ({\it the second row}), the number of external soft legs starts from three,  this then raises the questions about the number of effective terms needed  even at the lowest order. 

As discussed in Appendix \ref{sec:1lp_PT}, however, it is suffice to consider only the linear effective term at the one-loop order (equation \ref{eqn:linear_effterm}). 
This is because the power spectra constructed from these terms ({\it second row} in Figure.~ 
\ref{fig:hard_soft_1lp}) will be paired with a linear mode $\psi_b(-\vk)$, i.e.
$\left \langle \mS^{B^T (\Rmnum{1}, \Rmnum{2}, \Rmnum{3})}_{2\Lambda\tildeLambda, a} (\vk) 
 \psi_b(-\vk)\right \rangle$, 
and with the assumption of Gaussian initial condition, the only way is to pair modes 
$\vphi_{\vk-\vq}$ with $\vphi_{\vq-\vk}$, as highlighted by ellipses in Figure.~ (\ref{fig:hard_soft_1lp}). 
These effective terms $\mS^{B^T (\Rmnum{1}, \Rmnum{2}, \Rmnum{3})}_{2\Lambda\tildeLambda, a} (\vk) $ 
would then contribute to various soft-hard integral of $P_{13}$ and $P_{22}$.
Consequently, their effects could all be represented by some $k$-dependent factor multiplied 
by linear mode.

\begin{figure}
\includegraphics[width=1\textwidth]{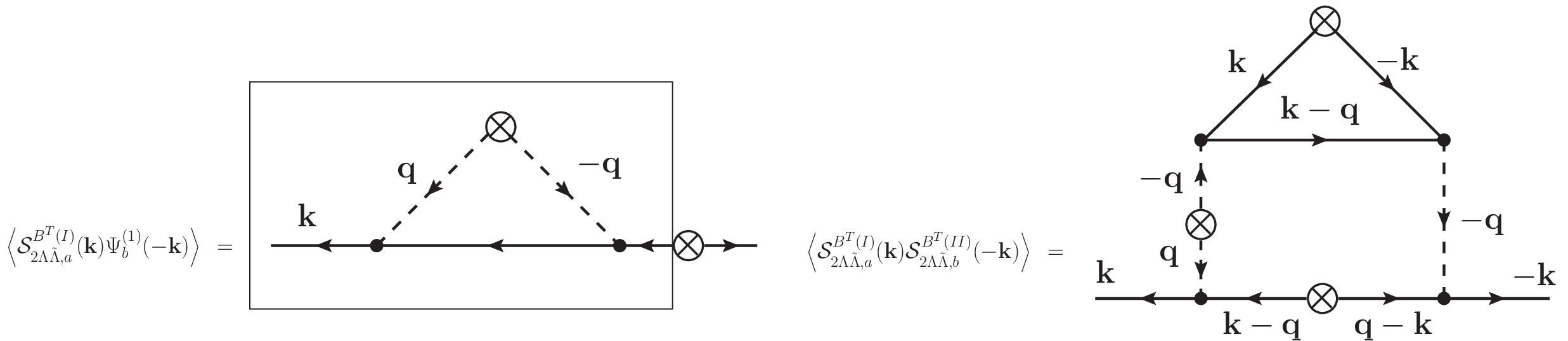}
\caption{\label{fig:pk_ct} Examples of 1PR ({\it left}) and 1PI ({\it right}) power spectra diagrams 
constructed from the hard-soft counter-terms $\mC^{B^T}_{2\Lambda\tildeLambda, a} (\vk)$. 
At one-loop order, it could be described by single parameters, i.e. $c^2$ term shown light 
box in the first diagram.
For higher-orders, this is not true any more. These non-local contributions would play an 
important role in constructing various SPT diagrams. 
}
\end{figure}

Particularly, for coefficient $ \mS^{B^T (\Rmnum{1}, \Rmnum{2})}_{UV, a} (\vk) $, 
where $UV=\{\tildeLambda\tildeLambda,  2\Lambda\tildeLambda\}$ denotes both hard-hard and
soft-hard couplings,  its ensemble average with $\psi_b(-\vk)$ give the full UV integral of the 
$P_{\delta, 13}(\vk)$, so the effective sound speed from this contribution could be expressed as
\begin{eqnarray}
\label{eqn:c13}
k^2 c^2_{\delta, 13, UV}(\vk) &=&   k^2 \int_{UV}
\frac{  2 \pi  q^2  P_{lin}(q) ~ dq d\mu }{21 \left(k^2+q^2\right)^2-4
   k^2 q^2   \mu^2}  
   \biggl [-21 k^4 \mu ^2+2 k^2  q^2 \left(38 \mu ^4-22 \mu ^2+5\right) \nonumber\\
&&  ~ + ~ q^4  \left(28 \mu ^4-59 \mu ^2+10\right) \biggr].
\end{eqnarray}
Explicit calculation demonstrates that $c^2_{\delta, 13}$ only mildly depends on $k$, 
so at the lowest order, it scales with $k^2$, consistent with the EFT argument.

Combining  Eq. (\ref{eqn:c13}) and the expansion series (\ref{eqn:Cexpansion_short}),  we could further expand the linear effective term $\mC^{(1)}_{UV,a}(\vk)$  as
\begin{eqnarray}
\label{eqn:cs_exp}
\mC^{(1)}_{UV,a}(\vk)  = \left [ \left( c^{(11)} \right)_{ab}^2  k^2 + \left( c^{(12)}\right)_{ab} ^2 k^4 + \cdots \right] \vpsi_b(\vk).  
\end{eqnarray}
{\revision
Readers who are familiar with EFTofLSS might be puzzled by the term $\left( c^{(12)} \right)_{ab}^2 k^4 \psi_b(\vk)$. In there, the $k^4$ coefficient that contributes to $P_{22}(k)$ is {\it not} proportional to $\vpsi$ and therefore is usually considered as stochastic. 
In our formalism,  $\left( c^{(12)}\right)_{ab} ^2 k^4  \vpsi_b(\vk)$ actually still contributes to $P_{22}(k)$.  
This is because that $ \mS^{B^T (\Rmnum{3})}_{UV, a} (\vk)$, which leads to the power spectrum contribution via $\left \langle \mS^{B^T (\Rmnum{3})}_{UV, a} (\vk) \psi_{b}(-\vk) \right \rangle $, is proportional to the inverse linear power spectra (equation \ref{eqn:Shh_BIII} and \ref{eqn:P22_hdhd} ). 
Consequently, at the one-loop order, we will have 
 \begin{eqnarray}
 k^4 c^2_{\delta, 22, UV} &=& \frac{k^4}{ P_{\rm L}(k) }  \int_{UV}   2 \pi  dq d\mu ~
    \frac{(7 k \mu + 3 q - 10 q \mu^2)^2 }{98 (k^2 + q^2 - 2 k q \mu)^2 }  
  ~   P_{\rm L}(q) P_{\rm L}(|\vk-\vq|) , 
 \end{eqnarray}
Practically, this $1/P(k)$ dependence would require a more careful treatment in measuring the coefficients.
}

For higher-loop orders, the linear term $\mC^{(1)}_a$ will not be enough, and we need to generally consider non-local terms. In the right panel of Figure.~ (\ref{fig:pk_ct}), we illustrate an example related to the time evolved coefficient  $\mS^{B^T}_{a} (\vk)$. 
Unlike the one-loop order (the left diagram), this one particle irreducible (1PI) contribution will {\it not} be accounted by a linear effective term. 
Therefore, this suggests that in order to have sufficient freedom to completely describe all UV effects, it is necessary to include all non-local terms. 

\subsection{Measurement of the Effective Terms}
\label{sec:effterm_measure}
In the above discussion, we have tried to expand the effective coefficients with standard mathematical tools (i.e. Gram-Charlier expansion). 
This is very different from the EFT framework, which expands in the number of local gradient operators $\nabla$ (or the power of $k$ in Fourier space). 
The Gram-Charlier series we used to estimate the conditional average (equation \ref{eqn:ct_def}) is an expansion of the degree of the non-Gaussianity, i.e. the cumulants of the field.  While there certainly exists a mapping between these two approaches like what we did in equation (\ref{eqn:cs_exp}), especially for those hard-hard coupling terms, we might benefit from performing both approaches. 

Undoubtedly, when measuring these effective terms, $\nabla$ (or $k$) expansion is beneficial as it has better control over the convergence of the perturbation calculation at a certain $k$ scale.  
But it might also conceal the internal structures among those counter-terms, making itself  vulnerable to the critique of overfitting.  In this regard, the cumulants expansion would serve as a self-calibration process, since these terms are simply some integral of high order statistics.

{\revision
Following the derivation in Sec.~(\ref{sec:Cexpansion}), once we measure the {\it non-linear} power spectrum and bispectrum etc. from N-body simulation, we could then calculate the lower order $\mC_a$ with Eq. (\ref{eqn:stpara_B_maintext}). 
This process might seem circular at first glance, but remember that our formalism recovers the full joint PDF $\mP(\vpsieff)$. Therefore, with all the measured $\mC_a$, we will still be able to predict all relevant statistics of $\vpsieff$ including topologies, covariance matrix, N-points PDF and higher order polyspectra etc.

Moreover, given the definition (Eq.~\ref{eqn:ct_def}) of $\mC_a \sim \int_{\rm UV} \langle \vpsi_{\vk_1} \vpsi_{\vk_2} | \vpsieff \rangle $,  there might be some interesting approaches to measure/calibrate these terms directly. 
One possible idea is to generate ensemble of N-body simulations with given soft-modes $\vpsieff$  \cite{Aragon_Calvo_2015}. Very much similar to the ``separate universe'' approach \cite{Li_2014,Wagner_2014}, one could then measure the average `response' of small-scale interaction to the soft modes set  $\vpsieff$. 
We will defer these numerical measurements for future studies. 
}

\subsection{$\Lambda$-dependence and Renormalization}
In the framework of effective field theory, it is critical to introduce the concept of regularization and 
renormalization. 
In our formalism, both large-scale coupling as well as effective contributions explicitly depend on
the cutoff scale $\Lambda$. 
However, assuming we are interested in a subset of large-scale modes, say $\vPsi_{\Gamma} \in \vPsi_{\Lambda}$, which does not really depend on $\Lambda$ as long as  $\Gamma$ is small.
Likewise, the evolution of the joint PDF of $\vPsi_{\Gamma}$ should not depend on $\Lambda$, 
because marginalizing over $\vPsi_{\tildeGamma}$ would be the same as marginalizing over
$\vPsi_{\tildeLambda}$ first then over $\vPsi_{\tildeGamma}/ \vPsi_{\tildeLambda}$, 
which is the complement set of $\vPsi_{\tildeLambda}$ with respect to $\vPsi_{\tildeGamma}$. 
This guarantees that our statistical effective dynamics of $\vPsi_{\Gamma}$ would
not depend on the cutoff scale $\Lambda$ either. 
Therefore, similar to the effective field theory, all our effective coefficients are composed of 
$\Lambda$-dependent and $\Lambda$-independent parts. The former appears simply to cancel 
out the $\Lambda$-dependence of the theory, leaving only the $\Lambda$-independent part.

\section{Conclusion and Discussion}

In this paper, we apply the so-called PDF-based method to study the evolution of joint PDF of coarse-grained cosmic field $\mP(\vpsi_{\Lambda}, \eta)$. 
We showed that the characteristic trajectory of this PDF evolution could serve as an effective dynamics because it recovers the exact statistics of the coarse-grained field. 
In this effective dynamics, small-scale physics are encoded in the effective terms expressed as the conditional average $\langle \vpsi \vpsi |\vpsi_{\Lambda} \rangle$. 
Naturally, these conditional averages could be expanded in the number of external large-scale modes, and we have showed that non-local terms are necessary for the complete recovery the statistics.
To estimate these effective terms, we applied the Gram-Charlier expansion, and demonstrated the agreement with SPT at one-loop order.  

As an effective theory, our formalism looks quite similar to the EFTofLSS at the linear order (in external fields), i.e. equation (\ref{eqn:linear_effterm}) and (\ref{eqn:cs_exp}), except that higher order $k$ dependence would also be captured by terms like equation (\ref{eqn:stpara_B}).
The distinction is due to different expansion series, i.e. cumulants expansion in our formula
versus the gradient expansion of EFT. 



Generally, this framework could also be applied to many other dynamical systems. 
For example, one interesting application would be the statistical evolution of biased tracers. 
Consider a smoothed field of both the number density fluctuation $\delta_t$ and the peculiar velocity 
$\vu_t$ of a particular type of tracer, denoted as $\vpsi_t$,  
we would like to understand the statistics of the large-scale modes of this field, i.e. 
$\vpsi_{t, \Lambda}$.
Following the same procedure, one could write down the kinetic equation of $\mP(\vpsi_{t, \Lambda}, \eta)$, 
and study its effective solution.
However, unlike the dark matter field, we are less certain about the fluid description of the biased tracer. 
For example, the number density $\delta_t$ does not necessarily conserve. Rather, due to merger, fragmentation 
and galaxy formation etc., one should expect an extra source term $j_t$ 
for the continuity equation (\ref{eqn:cdyn_eul}). So the effective dynamics would have a contribution like
$\langle j_t | \vpsi_{t, \Lambda} \rangle$.

\section*{Acknowledgments}
The author would like to thank for productive discussion with Diego Blas, Vincent Desjacques, Simon Foreman, Enrico Pajer, and Sergey Sibiryakov.

\appendix

\newcommand{\appsection}[1]{\let\oldthesection\thesection
\renewcommand{\thesection}{\oldthesection}
\section{#1}\let\thesection\oldthesection}

\section{Cosmic Dynamics}
\label{appsec:cdyn}


At the sub-horizon scale, the dynamics of the large-scale structure is well described by
the single-particle  phase space density $f(\vx, \vp, \tau)$ of the non-relativistic collisionless 
cold dark matter, which obeys the Vlasov equation \cite{BCGS12review} 
\begin{eqnarray}
\label{eqn:vlasov_eq}
\partial_{\tau} f + \frac{\vp}{ma}\cdot \nabla f 
- a m \nabla \Phi \cdot  {\bm \partial_p} f = 0, 
\end{eqnarray}
where $\tau$ here is the conformal time, $a(\tau)$ is the scale factor, $m$ is the mass of the dark matter, $\vp = am \dot{\vx} $ is the momentum of the particle, and $\Phi$ is gravitational potential,  determined by the Poisson equation
\begin{eqnarray}
\label{eqn:poisson_eq}
\nabla^2 \Phi = 4\pi G \bar{\rho} a^2 \delta. 
\end{eqnarray}
Here $G$ is the gravitational constant,  $\delta=\rho/\bar{\rho} - 1$ is the density contrast,
and $\bar{\rho}(\tau)$ the average density. 
To avoid solving this $(6+1)$-dimensional non-linear partial differential equation, one instead
takes the moments of $\vp$, and obtain a hierarchy of differential equations. The first two 
are continuity and Euler equation
\begin{eqnarray}
\label{eqn:cdyn_eul}
\partial_{\tau} \delta + \nabla_i  \left [ (1+\delta) u_i \right] &=& 0,  \nonumber \\
\partial_{\tau} u_i + (u_j \nabla_j) u_i + \mH u_i &=& - \nabla_i \Phi - \pi_i,  
\end{eqnarray}
where $\vu$ is the peculiar velocity, and $\mH(\tau) = d\ln a/d\tau$.  
Clearly, this equation is not solvable, since we have also included the second
moment of $f(\vx,\vp, \tau)$, i.e. the velocity dispersion $\sigma^u_{ij}$
\begin{eqnarray}
\label{ean:v_disp}
\rho \sigma^u_{ij} &=&  \int d^3\vp ~ \frac{p_i p_j} {a^2 m^2} 
     f(\vx, \vp, \tau)  - \rho u_i u_j , 
\end{eqnarray}
via the definition $\pi_i =  (\nabla_j \rho \sigma^u_{ij} ) / \rho$.
In the Fourier space, after decomposing $\vu$ into the divergent $\theta$ and vorticity $\vomega$, the evolution equation (\ref{eqn:cdyn_eul}) could be expressed as  
\begin{eqnarray}
\label{eqn:cosdyn_all}
\partial_{\eta} \delta(\vk) - \theta(\vk) &=& \alpha(\vk_1, \vk_2) \theta(\vk_1) \delta(\vk_2) 
 +~ \alpha^{\omega}_{i}(\vk_1, \vk_2)\omega_i(\vk_1) \delta(\vk_2)  \qquad\qquad \quad~ ~ \nonumber \\
\partial_{\eta} \theta(\vk) + (g-1) \theta(\vk) - g \delta(\vk) &=&  \beta(\vk_1, \vk_2) \theta(\vk_1) \theta(\vk_2)   
 + \beta^{\omega}_{i}(\vk_1,\vk_2)  \omega_i(\vk_1) \theta(\vk_2)  \nonumber \\
&& ~ + \beta^{\omega^2}_{ij}(\vk_1,\vk_2) 
  \times \omega_i(\vk_1) \omega_j(\vk_2) + \pi_{\theta}(\vk)   \nonumber \\
\partial_{\eta} \omega_i(\vk) + (g-1) \omega_i(\vk)& =& \kappa^{\omega}_{ij} (\vk_1, \vk_2) \omega_j(\vk_1)
\theta(\vk_2)  + \kappa^{\omega^2}_{ijk} \omega_{j}(\vk_1) \omega_k(\vk_2) + \pi_{\omega,i}(\vk) . \qquad  \qquad
\end{eqnarray}
Here $g=3\Omega_m/(2f^2) \approx 3/2$, and we have introduced
the new time variable $\eta$ so that $d\eta=d \ln D(\tau)$, where $D(\tau)$ is the linear
growth rate. We have also rescaled the velocity $u_i\to -\mH f u_i$, 
and $\pi_i \to (\mH f)^2 \pi_i $,  
and  the vector $\vpi$ is decomposed into longitudinal $\pi_{\theta} = \nabla \cdot \vpi$ and transverse
$\vpi_{\omega} = \nabla \times \vpi$ parts. 
Besides the standard kernel, here, we have also introduced extra coupling terms among 
the vorticity $\vomega$ and $\delta$ or $\theta$, i.e. $\alpha^{\omega}, \beta^{\omega}, 
\beta^{\omega^2}$ and $\kappa^{\omega}, \kappa^{\omega^2}$.
Please read Appendix \ref{appsec:cdyn} for the definition of these kernels.

Usually, to close the system, one has to drop the velocity dispersion term $\vpi$, which is justifiable only before the emergence of the shell-crossing. 
For the same reason, the vorticity is also neglected because any primordial rotation would have decayed away without the source term. In the era of precision cosmology, this has created a situation that enormous efforts have been made to improve the prediction accuracy of the LSS clustering based on a set of approximate equations. As demonstrated by other authors (e.g. \cite{PS09}), this dust model will introduce at least $\sim 1\%$ error in the BAO regime,  comparable to the statistical uncertainty of the next generation LSS survey.  

Standard kernels:
\begin{eqnarray}
 \alpha(\vk_1, \vk_2) &=&  \delta_D(\vk-\vk_{12}) \frac{ (\vk_{12}\cdot \vk_1)}{k_1^2}, \nonumber\\
\beta(\vk_1, \vk_2) &= & \delta_D(\vk-\vk_{12}) \frac{ k_{12}^2 (\vk_1\cdot \vk_2)}{2 k_1^2 k_2^2},  
\end{eqnarray}
where $\vk_{12}=\vk_1+\vk_2$. The extra kernels are 
\begin{eqnarray}
\alpha^{\omega}_{i}(\vk_1, \vk_2) &=& \delta_D(\vk-\vk_{12})  \frac{  (\vk_1 \times \vk_2)_i }{k_1^2}  \nonumber \\
\beta^{\omega}_{i} (\vk_1,\vk_2) &=&  \delta_D(\vk-\vk_{12})  \frac{ 2 \vk_1\cdot \vk_2 +  k_2^2}{k_1^2 k_2^2 }  
   (\vk_1 \times \vk_2)_i  \nonumber \\
 \beta^{\omega^2}_{ij} (\vk_1,\vk_2) &=& \delta_D(\vk-\vk_{12}) \frac{     (\vk_2 \times \vk_1)_i  (\vk_1 \times \vk_2)_j }
 {k_1^2 k_2^2} \nonumber\\
 \kappa^{\omega}_{ij}(\vk_1, \vk_2)&=& \delta_D(\vk-\vk_{12}) \frac{(\vk_{12} \cdot \vk_2) 
 \delta_{ij}  - k_{2i} k_{2j} }{k_2^2}  \nonumber \\
 \kappa^{\omega^2}_{ijk}(\vk_1, \vk_2) &=& \delta_D(\vk-\vk_{12}) \frac{  \epsilon_{ilk} k_{2j} k_{2l} -
 (\vk_1 \times \vk_2)_k \delta_{ij} } {k_2^2}, \nonumber \\
\end{eqnarray}
$\epsilon_{ijk}$ is the Levi-Civita symbol.

We are seeking a statistical closure of this dynamics. Since we are not particularly interested in the evolution of vorticity itself, we will only consider it as some extra source term similar to $\vpi$. 
Defining the dynamical vector for matter field as $\vpsi(\vk)=\{ \delta(\vk), \theta(\vk) \}$, 
one could express equation (\ref{eqn:cdyn_eul}) in a compact form \cite{BCGS12review}
\begin{eqnarray}
\label{eqn:fourier_dyn}
\hat{\mL}_{ab}  \psi_b (\vk)  
&=& \gamma_{abc}(\vk_1, \vk_2) \psi_b(\vk_1) \psi_c(\vk_2)  
+ \gamma^{\omega}_{aib} (\vk_1, \vk_2) 
\omega_i(\vk_1) \psi_b(\vk_2)   + \gamma^{\omega^2}_{aij} (\vk_1, \vk_2) 
 \omega_i(\vk_1) \omega_j(\vk_2)  \nonumber\\
 && + \pi_a(\vk) , 
\end{eqnarray}
where index $a,b,c \in \{ 1, 2\}$, and $i, j$ are spatial indices of vectors. 
The linear operator $\hat{\mL}_{ab} $ is defined as
\begin{eqnarray}
	\label{eqn:Lab}
\hat{\mL}_{ab} = \partial_{\eta} \delta_{ab} + \Omega_{ab},  
\end{eqnarray}
and the coefficient matrix $\Omega_{ab}$ equals 
\begin{eqnarray}
\label{eqn:Omegaab}
\Omega_{ab} = 
\begin{bmatrix}
0 & -1 \\
- g 
& g-1 
\end{bmatrix}
\approx
\begin{bmatrix}
0 & -1 \\
-3/2 & 1/2
\end{bmatrix}. 
\end{eqnarray}
The mode coupling vertex $\gamma_{abs}(\vk_1, \vk_2)$ is nonzero only at
\begin{eqnarray}
	\gamma_{121}(\vk_1, \vk_2) =  \alpha(\vk_1, \vk_2)/2,  ~~
	\gamma_{112}(\vk_1, \vk_2) =  \alpha(\vk_2, \vk_1)/2, ~~
	\gamma_{222}(\vk_1, \vk_2) =  \beta(\vk_1, \vk_2), ~~ \nonumber\\
	\gamma^{\omega}_{1i1} (\vk_1, \vk_2) =  \alpha^{\omega}_i (\vk_1, \vk_2) , ~~
 	\gamma^{\omega}_{2i2} (\vk_1, \vk_2) = \beta^{\omega}_i (\vk_1, \vk_2), ~~
   \gamma^{\omega^2}_{2ij} (\vk_1, \vk_2) =  \beta^{\omega^2}_{ij} (\vk_1, \vk_2)  . ~~
\end{eqnarray} 
And finally we define $\pi_a(\vk, \eta)=\left \{0,   \pi_{\theta}(\vk)  \right \}$.

In the standard pressureless perfect fluid (or dust model), which we neglect both $\vpi$ and 
$\vomega$ terms, one could formally express the nonlinear solution as \cite{BCGS12review,CS06a,CS06b}
\begin{eqnarray}
\label{eqn:form_solution_dm}
\psi_a(\vk, \eta) &=& g_{ab}(\eta)  \phi_b(\vk) + \int_0^{\eta} ds ~g_{ab}(\eta-s) \gamma_{bcd}(\vk_1,\vk_2) 
 \psi_c(\vk_1, s)   \psi_d(\vk_2, s)  
\end{eqnarray}
where $\phi_a(\vk)=\psi_a(\vk, \eta_{\ini})$ is initial condition, and the linear propagator $g_{ab}(\eta)$ is 
\begin{eqnarray}
g_{ab} (\eta) = \frac{e^{\eta}}{5}
\begin{bmatrix}
3 & 2 \\
3 & 2 
\end{bmatrix}
-  \frac{e^{-3\eta/2}}{5}
\begin{bmatrix}
-2 & 2 \\
3 & -3 
\end{bmatrix} .
\end{eqnarray}
The growing initial condition is $\phi_a \propto [ 1, 1 ]$, 
and one notices that $g_{\alpha\beta}$ is invertible as long as we keep both growing
and decaying mode. 
This formal solution leads to simple diagram representation of $\psi_a(\vk, \eta)$ \citep{CS06a,CS06b},
which we will adopt in the following of the paper. 
The standard perturbation series could be expanded as $\psi_a(\vk, \eta)=\sum_n \psi_a^{(n)}(\vk, \eta)$,
\begin{eqnarray}
\psi_a^{(n)} (\vk; \eta)&=& \int_{\vk_{1\cdots n}}  \mF^{(n)}_{a a_1 \cdots a_n}  (\vk_1, \cdots \vk_n; \eta )
\phi_{a_1}(\vk_1) \cdots \phi_{a_n}(\vk_n), 
\end{eqnarray}
where $ \mF^{(n)}_{a a_1 \cdots a_n}$ is the SPT kernel.
Finally, the power spectrum is defined as
\begin{eqnarray}
\langle  \psi_{a}(\vk) \psi_b(\vk^{\pri})   \rangle_c = \delta_D(\vk+\vk^{\pri}) P_{ab} (\vk), 
\end{eqnarray}
where  subscript $_c$ denotes the connected part of the average.

\appsection{Perturbative Expansion of Conditional Expectation in the Weakly Non-Gaussian Region}
\label{app:condexp}
We have to estimate the conditional expectation of the form 
$\langle x_1 | \vY \rangle$ and $ \langle x_1 x_2 | \vY \rangle$, where $\vY$. 
By definition, this would be expressed as
\begin{eqnarray}
	\langle x_1 \cdots x_n | \vY \rangle \mP(\vY) = \int d\vX ~ (x_1 \cdots x_n) ~ \mP(\vGamma), 
\end{eqnarray}
where $\mP(\vGamma) = \mP(\vX, \vY)$. Following the derivation in \cite{WS16}, we have for Gaussian variables, 
\begin{eqnarray}
	\label{eqn:condavg_gauss}
	\langle x_1  | \vY \rangle_G &=& \xi^{x_1 Y}_{\alpha} \left(\xi^Y \right)^{-1}_{\alpha\beta} Y_{\beta}  \nonumber\\
	\langle x_1  x_2 | \vY \rangle_G& =& \xi^{x_1 x_2} + \xi^{x_1 Y}_{\alpha}   \xi^{x_2 Y}_{\beta} 
	\biggl [  \left(\xi^Y\right)^{-1}_{\alpha\gamma}  \left(\xi^Y \right )^{-1}_{\beta\delta} Y_{\delta}Y_{\delta}
	- \left(\xi^Y \right)^{-1}_{\alpha\beta}  \biggr ] 
\end{eqnarray}

Using the cumulants expansion theorem, we could further expand to the weakly non-Gaussian field
\begin{eqnarray}
	\label{eqn:cond_ave_hs}
	\langle x_1 | \vY \rangle & \approx& \langle x_1 | \vY \rangle_G +  \frac{1}{2} \mP_G^{-1}(\vY) 
	\biggl( \partial^{2Y}_{\alpha\beta} \mP_G(\vY) \biggr)
	\biggl [  \xi^{x_1 YY}_{\alpha\beta} 
	- \left( \partial^Y_{\gamma}\langle x_1 | \vY \rangle_G \right) \biggr] \nonumber \\
	& = &  \langle x_1 | \vY \rangle_G +  \frac{1}{2} \biggl [ \xi^{x_1YY}_{\alpha\beta}  -
	\xi^{YYY}_{\alpha\beta\gamma} \xi^{x_1 Y}_{\kappa} \left( \xi^Y \right)^{-1}_{\kappa\gamma}   \biggr] 
	\left[ \left( \xi^Y \right)^{-1}_{\alpha\lambda}  \left( \xi^Y \right)^{-1}_{\beta\tau} Y_{\lambda}
	Y_{\tau} -  \left( \xi^Y \right)^{-1}_{\alpha\beta}    \right]
\end{eqnarray}

On the other hand, we have
\begin{eqnarray}
	\label{eqn:cond_ave_hh}
	\langle x_1 x_2 | \vY \rangle  &\approx& \langle x_1 x_2 | \vY \rangle_G \biggl [1 + \frac{1}{3!}\xi^Y_{\alpha\beta\gamma}
	\mP_G^{-1} (\vY) \left ( \partial^{3Y}_{\alpha\beta\gamma} \mP_G(\vY) \right) \biggr] 
	- \frac{1}{3!} \mP_G^{-1} (\vY) \int d\vX (x_1 x_2) \xi^{\Gamma}_{\alpha\beta\gamma} 
	\left ( \partial^{3\Gamma}_{\alpha\beta\gamma} \mP_G(\vGamma)  \right) \nonumber \\
	&=& \langle x_1 x_2 | \vY \rangle_G + \xi^{x_1 x_2 Y}_{\alpha} \xiYinv_{\alpha\beta} Y_{\beta}
	- \frac{1}{2} \biggl[  \xi^{x_1YY}_{\alpha\beta}  
	\xi^{x_2Y}_{\gamma} + \xi^{x_2YY}_{\alpha\beta} \xi^{x_1 Y}_{\gamma} \biggr]  
	\biggl[  \xiYinv_{\gamma\delta} \xiYinv_{\alpha\gamma} \xiYinv_{\beta\tau} \nonumber \\
	&& \times Y_{\delta} Y_{\gamma} Y_{\tau}
	- \biggl( \xiYinv_{\alpha\gamma}    \xiYinv_{\beta\delta}  +   \xiYinv_{\beta\gamma}  \xiYinv_{\alpha\delta} 
	+   \xiYinv_{\alpha\beta}  \xiYinv_{\gamma\delta}  \biggr) Y_{\delta}   \biggr]  .
\end{eqnarray}

\appsection{One-loop Calculation of Effective Coefficients}
\label{sec:1lp_PT}

Although we have to seek the help of numerical simulation eventually, one still gain valuable insight by 
some analytic calculations.  
From the definition equation (\ref{eqn:condav_def}), we first notice that a Gaussian $\mP(\vGamma)$
will not produce any non-trivial result. 
However, by applying the so-called Gram-Charlier expansion of the non-Gaussian PDF $\mP(\vPsi_{\Lambda})$
and $\mP(\vGamma)$, we would be able to perturbatively calculate the conditional average. 
In general, this expansion could be written as \cite{C67,JWACB95,Amen96,BM98}
\begin{eqnarray}
\label{eqn:GC_expansion}
\mP(\vGamma) = \mP_G(\vGamma) \left [ 1+ \sum_{n\ge 3} \frac{1}{n!} \langle \vGamma^n \rangle_{GC}
\otimes_T  {\bm H}_n (\vGamma)   \right], 
\end{eqnarray}
where $\otimes_T$ denotes the inner produce of two tensors, $\mP_G$ is the Gaussian part of $\mP$, 
and the Hermite tensor ${\bm H}_n$ is 
defined as ${\bm H}_n(\vGamma) = (-1)^n \mP_G^{-1}(\vGamma) \partial^n \mP_G(\vGamma) /\partial^n \vGamma$. 
The conditional average could then be expressed in terms of Gram-Charlier coefficients 
$\langle \vGamma^n \rangle_{GC}$ which then related to the cumulants of various modes, 
i.e. the polyspectra here.

Since by construction, our effective solution recovers the statistics of the real system,  it would be interesting to examine whether this is also true at the perturbative level.  Naively, since the formalism applies regardless of the initial condition, one would expect 
this will be the case.  In the rest of the section, we will expand these counter-terms to the one-loop order, i.e. up to the
bispectra in equation (\ref{eqn:GC_expansion}). 
To proceed, however, one first notices that the condition average would be different for   the {\it hard-hard} ($\tildeLambda\tildeLambda$)  and {\it soft-hard} ($\Lambda\tildeLambda, \tildeLambda\Lambda$) couplings.

Equation (\ref{eqn:ct_taylor_exp}) expands the $\mC_a(\vk, \vPsi_{\Lambda}, \eta)$ as a function of 
$\Psi_a(\vk, \eta)$ at  some later time $\eta$. 
Following the spirit of standard perturbation theory, we would also like to expand $\mC_a(\vk, \eta)$ in terms of
initial  field $\vphi_a$ so that
\begin{eqnarray}
\mC_a(\vk, \vPsi_{\Lambda}; \eta)=  \sum_n
\mC^{(n)}_a(\vk, \vPsi_{\Lambda}; \eta). 
\end{eqnarray}

\begin{figure}
	\includegraphics[width=1\textwidth]{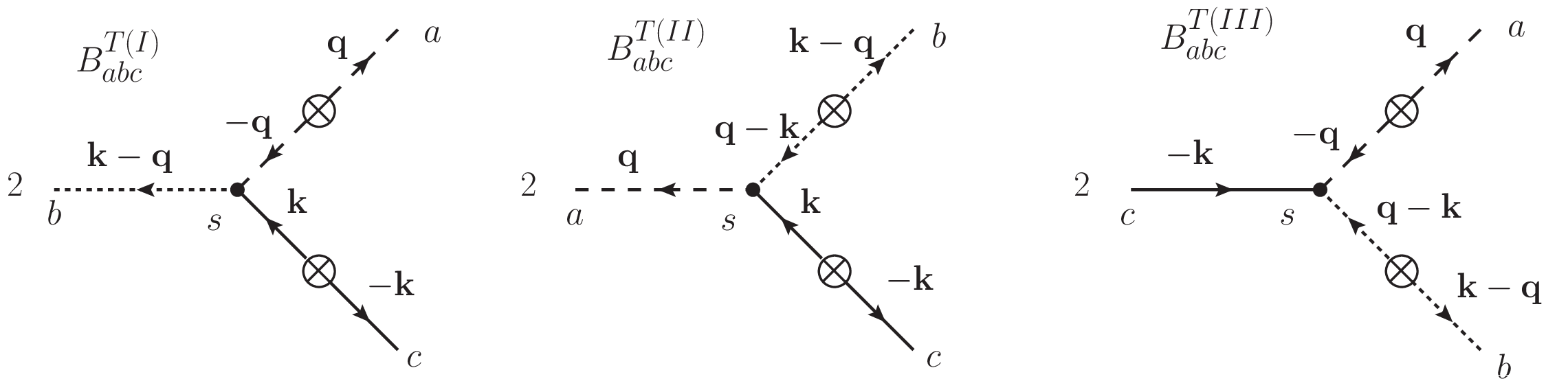}
	\caption{\label{fig:tree_bispectrum}
		Tree order bispectra that will be used for perturbative calculation of the effective 
		counter-terms $\mC_a(\vk, \vPsi_{\Lambda})$. 
		Solid lines denote soft modes $q<\Lambda$, and long-dashed lines indicate hard modes, i.e. $q>\Lambda$, 
		while short-dashed lines could be either hard or soft mode.
		A circle with a cross inside denotes the initial power spectrum. 
	}
\end{figure}

\subsection{Hard-hard Coupling}
The hard-hard coupling involves the ensemble average of two random small-scale modes, 
conditional on large-scale modes, i.e. $\langle \psi_{b,\vk_1} \psi_{c, \vk_2} | \vPsi_{\Lambda} \rangle$. 
Denoting $x_{1,2}$ as modes $\psi_{b,\vk_1}$ or $\psi_{c, \vk_2}$ respectively, and
$\vY = \vPsi_{\Lambda}$,  the Gaussian part of this average equals  
\begin{eqnarray}
\langle x_1  x_2 | \vY \rangle_G& =& \xi^{x_1 x_2} + \xi^{x_1 Y}_{\alpha}   \xi^{x_2 Y}_{\beta} 
\biggl [  \left(\xi^Y\right)^{-1}_{\alpha\gamma}  \left(\xi^Y \right )^{-1}_{\beta\delta} Y_{\delta}Y_{\delta} 
 - \left(\xi^Y \right)^{-1}_{\alpha\beta}  \biggr ] , 
\end{eqnarray}
where $ \xi^{x_1 x_2}$ is the correlation function between $x_1$ and $x_2$, and $\xi^{x_1 Y}_{\alpha} $
the correlation between $x_1$ and $\alpha$-th element of $\vY$. Similarly, $\xi^Y_{\alpha\beta}$ is the
correlation between $\alpha$-th and $\beta$-th elements of $\vY$, and $ \left(\xi^Y\right)^{-1}_{\alpha\beta}$
is the inverse. 
Since by the definition of $\mC_{\alpha}(\vk, \vpsi_{\Lambda}, \eta)$ (equation \ref{eqn:ct_def}), 
$\vk_1+\vk_2 = \vk \ne 0$, the first term vanishes because of the statistical translational invariance. 
Similarly, the second contribution is also zero since $x_{1,2}$ and modes in $\vY$ belong to different scales. 
The next leading order, which we present its full formula in equation (\ref{eqn:cond_ave_hh}), has only 
one non-vanishing contribution
\begin{eqnarray}
\langle x_1 x_2 | \vY\rangle =  \xi^{x_1 x_2 Y}_{\alpha} \xiYinv_{\alpha\beta} Y_{\beta}, 
\end{eqnarray}
where $ \xi^{x_1 x_2 Y}_{\alpha} = \langle x_1 x_2 Y_{\alpha} \rangle_c$ is the bispectra of 
$x_1$, $x_2$ and $Y_{\alpha}$.

Therefore, to  the first order, the hard-hard-bispectra parts of the effective coefficients 
$\mC^{B}_{\tildeLambda\tildeLambda,a}(\vk, \eta)$  is expressed as an integral of
the bispectrum with the kernel $\gamma_{abc}$, 
\begin{eqnarray}
\label{eqn:stpara_B}
\mC^{B}_{\tildeLambda\tildeLambda,a}(\vk, \eta) &=& \int_{\tildeLambda\tildeLambda} 
d\vq ~ \gamma_{abc} (\vq, \vk-\vq)  B_{bcd} (\vq, \vk-\vq, -\vk; \eta)  
 P^{-1}_{de}(k, \eta)  \psi_e(\vk, \eta), 
\end{eqnarray}
where  the integral is over the region where both $q$ and 
$|\vk-\vq|$ are greater than $\Lambda$,  and all the quantities here are evaluated at time $\eta$. 
Here the bispectrum $B(\vq, \vk-\vq, -\vk)$  should be fully non-linear,
as well as the power spectra $P_{de}(k, s)$ and its inverse. 
This would alleviate the problem of singular IC since even at some early stage, a tiny amount of 
nonlinearity would be able to render the $P^{-1}(k, s)$ mathematically well-defined.

At the lowest order, we can take the tree-level bispectrum $B^T \propto P_{lin}^2$, where $P_{lin}$  is the linear power spectrum, and this is already sufficient to produce the one-loop order of  $\mC_a^{(1)}(\vk, \vPsi_{\Lambda}, \eta)$. 
\begin{eqnarray}
	\label{eqn:Btree_sum}
	B^{T}_{bcd} (\vq, \vk-\vq, -\vk) = B^{T (\Rmnum{1})}_{bcd} + B^{T (\Rmnum{2})}_{bcd}  + 
	B^{T (\Rmnum{3})}_{bcd}. 
\end{eqnarray}
In the standard formula, they are defined as
\begin{eqnarray}
	\label{eqn:Btree_I}
	B^{T (\Rmnum{1})}_{bcd} (\vq, \vk-\vq, -\vk) &=& 2\int_0^{\eta} ds ~ g_{c\alpha} (\eta-s) 
	\gamma_{\alpha\beta\gamma}(-\vq, \vk) 
	\biggl [ g_{\beta\delta}(s) g_{b\tau}(\eta)  P^{\ini}_{\delta\tau}(q) \biggr]  \nonumber \\
	&& ~ \times  ~ \biggl [ g_{\gamma\lambda}(s) 
	g_{d\epsilon}(\eta) P^{\ini}_{\lambda\epsilon}(k) \biggr],  
\end{eqnarray}
where 
\begin{eqnarray}
	\label{eqn:Btree_cyc}
	B^{T (\Rmnum{2})}_{bcd}(\vq, \vk-\vq, -\vk)  =  B^{T (\Rmnum{1})}_{cbd}(\vk-\vq, \vq, -\vk), 
	B^{T (\Rmnum{3})}_{bcd}(\vq, \vk-\vq, -\vk) =  B^{T (\Rmnum{1})}_{cdb}(\vk-\vq, -\vk, \vq)  ~~ ~
\end{eqnarray}
The diagrams of these three contributions are shown in Figure.~ (\ref{fig:tree_bispectrum}).  In the rest of the paper, we will always assume $\vk$ ($\vq$) denotes the wavenumber of some soft (hard)  mode.

\begin{figure*}
	\includegraphics[width=0.95\textwidth]{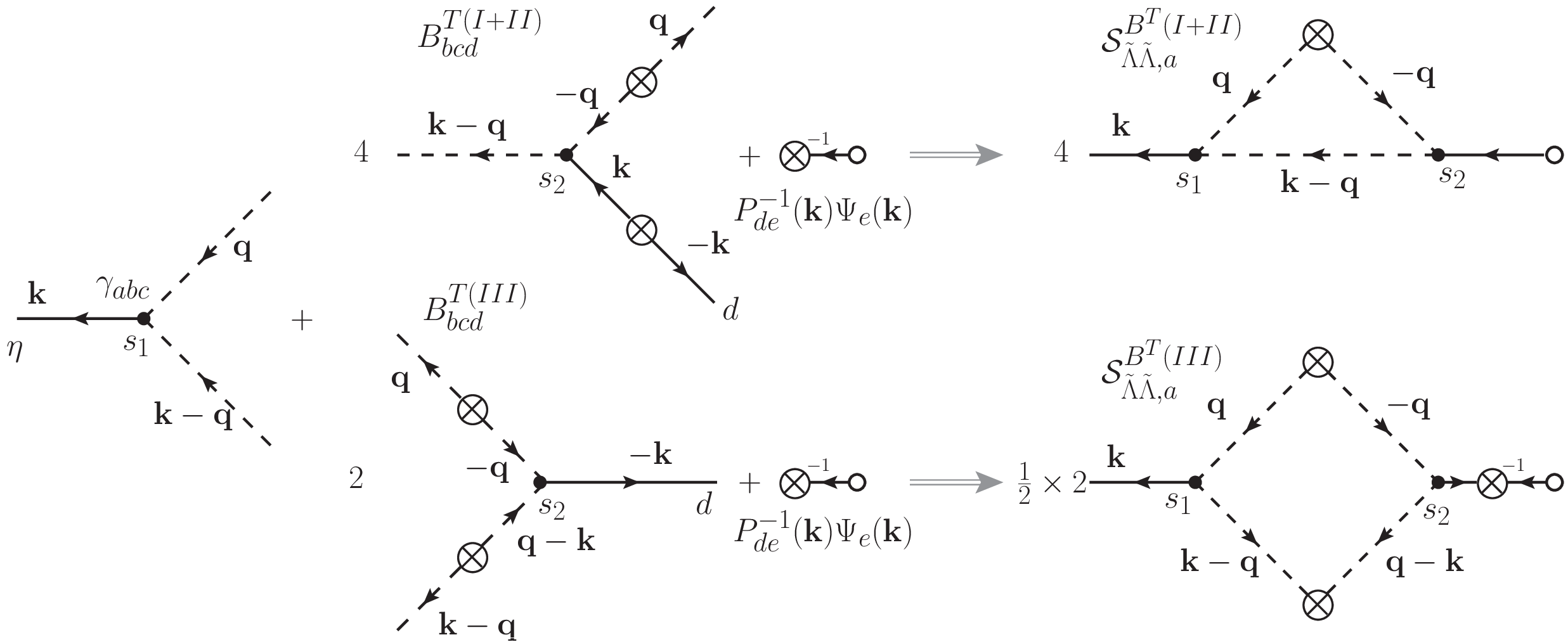}
	\caption{\label{fig:P1l_bi}
		Schematics that demonstrate the Feynman rules for constructing the evolved hard-hard effective terms 
		$\mS_{\tildeLambda\tildeLambda,a}(\vk, \vPsi_{\Lambda}, \eta)$ at the one-loop order.  
		Specifically, equation (\ref{eqn:ct_def}) and (\ref{eqn:stpara_B}) indicates that one only need to 
		connect the symmetric perturbative kernel $\gamma_{abc}(\vq, \vk-\vq)$ together with conditional 
		average terms, which in one-loop order is the tree-level bispectra 
		$B^{T(\Rmnum{1},\Rmnum{2},\Rmnum{3})}$ and the inverse power spectrum $P_{de}^{-1}(\vk)$. 
		The dashed-lines denote the hard modes, while solid lines represent soft modes.
		The numerical factor $1/2$ in front of the last diagram is caused by releasing the causal relation $s_2\le s_1$ 
		assumed initially. 
		From the diagram representation of these two contributions, we could already see that they would recover the hard-hard part of the one-loop power spectrum exactly simply by taking the average with some linear solution  $\vpsi^{(1)}(\vk, \eta)$.
	}
\end{figure*}

Substituting the equation (\ref{eqn:Btree_I}) back into equation (\ref{eqn:stpara_B}), we  then have three separate contributions.  
Since the kernel $\gamma_{abc}(\vq, \vk-\vq)$ is symmetric with respect to $\vq$ and $\vk-\vq$,  the contribution from $B^{T (\Rmnum{1})}$ and $B^{T (\Rmnum{2})}$ would be identical.  
Therefore, the effective coefficient reads  
\begin{eqnarray}
\label{eqn:C_BT_hh_12}
 \mC^{B^{T} (\Rmnum{1}+\Rmnum{2})}_{\tildeLambda\tildeLambda, a}(\vk, \eta) &=& 4 \int_{\tildeLambda\tildeLambda} d\vq 
 ~ \gamma_{abc}(\vq, \vk-\vq)  \int_0^{s} ds ~ g_{c\alpha}(\eta-s) 
  \gamma_{\alpha\beta\gamma} (-\vq, \vk) \biggl[ g_{\beta\delta}(s)g_{b\tau}(\eta) 
 P^{\ini}_{\delta\tau}(q) \biggr]  \nonumber \\ 
 &&  \times~ \psi_{\gamma} (\vk, s_2). 
\end{eqnarray} 
Here we have already applied the identity that 
\begin{eqnarray}
\label{eqn:Pinv_identity}
 \delta^K_{\gamma\tau} &=& g_{\alpha\beta}(s) P^{\ini}_{\beta\gamma}(k) P^{-1}_{\alpha \delta} (k, s) g_{\delta\tau} (s)  \nonumber \\
& =   & g_{\alpha\beta}(s) P^{\ini}_{\beta\mu}(k) g_{\mu\nu}(s) g^{-1}_{\nu \gamma}(s) 
P^{-1}_{\alpha \delta} (k, s) g_{\delta\tau} (s)  , 
\end{eqnarray}
assuming $P_{\alpha\delta}(k, s)$ is not singular, and we have only kept the linear part of its inversion
$P^{-1}_{lin}(k)$. 
One could further write down the time-evolved coefficient 
$\mS^{B^T}_{\tildeLambda\tildeLambda, a}(\vk, \eta)$, whose Feynman diagram is presented in 
Figure.~ (\ref{fig:P1l_bi}). 
Particularly, in the same figure, we demonstrate the general rule for constructing these `counter-terms', 
which is simply paring the symmetric PT kernel $\gamma_{abc}(\vq, \vk-\vq)$ together with various contribution from the conditional average terms, i.e.  $B^{T(\Rmnum{1},\Rmnum{2})}$ and $P^{-1}_{lin} \vpsi$  here.

From the diagram, it is obvious that one would recover the one-loop power spectrum 
$P^{13}(k)$ with the contribution from 
$\mS^{B^T (\Rmnum{1}+\Rmnum{2})}_{\tildeLambda\tildeLambda, a}(\vk, \eta)$  
\begin{eqnarray}
\label{eqn:P13_lam}
 P^{(13)}_{\tildeLambda\tildeLambda, ab} (\vk, \eta) &=& 2
\left \langle   \mS^{B^T (\Rmnum{1}+\Rmnum{2})}_{\tildeLambda\tildeLambda, a}(\vk, \eta)
\psi^{(1)}_b (-\vk, \eta)  \right \rangle  \nonumber \\
&=& 6 P^{\ini}(k) \int_{\tildeLambda\tildeLambda}   d\vq~  
\mF_{a}^{(3)}(\vk, \vq, -\vq; \eta) P^{\ini}(q)  . 
\end{eqnarray}
Again, here we are only integrating over the Fourier region $q>\Lambda$ and 
$|\vk-\vq| > \Lambda$, which corresponds to the region $q>\Lambda$
and $\mu < (k^2+q^2-\Lambda^2)/(2kq)$, where $\mu$ is the cosine of the angle between 
$\vk$ and $\vq$. 
This does seem a bit odd in SPT formula since there is only one integral in equation 
(\ref{eqn:P13_lam}), and $\vk-\vq$ does not even appear in the definition of $\mF^{(3)}$,
but the meaning is clear from the time-evolved representation like equation 
(\ref{eqn:C_BT_hh_12}).

Equation (\ref{eqn:P13_lam}) helps us to derive a much simpler expression for 
$\mS^{B^T (\Rmnum{1}+\Rmnum{2})}_{\tildeLambda\tildeLambda, a}(\vk,\eta)$ with the
kernel of the standard perturbation theory 
\begin{eqnarray}
\label{eqn:Sbt1_hh_kern}
 \mS^{B^T (\Rmnum{1}+\Rmnum{2})}_{\tildeLambda\tildeLambda, a}(\vk, \eta)
 &=& 3 \int_{\tildeLambda\tildeLambda}   d\vq~  \mF_{ac}^{(3)}(\vk, \vq, -\vq; \eta)
 P^{\ini}(q) \phi_c(\vk) , 
\end{eqnarray}
where $\mF_{ac}^{(3)}(\vk, \vq, -\vq)= \mF_{acde}^{(3)}(\vk, \vq, -\vq) u_d u_e$. 
This further suggests that the effective coefficient should be
\begin{eqnarray}
 \mC^{B^T (\Rmnum{1}+\Rmnum{2})}_{\tildeLambda\tildeLambda, a}(\vk, s) &=& 
 6 \int_{\tildeLambda\tildeLambda}   d\vq~  \mF_{ab}^{(3)}(\vk, \vq, -\vq; s) P^{\ini}(q)\phi_b(\vk)  \nonumber\\
&=&  6 \int_{\tildeLambda\tildeLambda}   d\vq~  \mF_{a}^{(3)}(\vk, \vq, -\vq; s) P^{\ini}(q) \delta_{\ini}(\vk).
\end{eqnarray}

Furthermore, we then substitute the last $B^{T (\Rmnum{3})}$ into the equation (\ref{eqn:stpara_B}), 
the coefficient reads
\begin{eqnarray}
\label{eqn:Shh_BIII}
 \mC^{B^T (\Rmnum{3})}_{\tildeLambda\tildeLambda, a}(\vk, \eta) &=& 2 \int_{\tildeLambda\tildeLambda}
 d\vq ~ \gamma_{abc}(\vq, \vk-\vq)   \int_0^{\eta} ds ~g_{d\alpha}(\eta-s)  
 \gamma_{\alpha\beta\gamma}(\vq-\vk, -\vq)  \biggl [ g_{\beta\delta}(s)
 g_{c\tau}(\eta) \nonumber \\
 && \times P^{\ini}_{\delta\tau}(|\vk-\vq|) \biggr]   
 \biggl [ g_{\gamma\lambda}(s) g_{b\epsilon}(\eta) P^{\ini}_{\lambda\epsilon}(q) \biggr]
 P^{-1}_{de} (k, s)  \psi_e (\vk, s)
\end{eqnarray} 
Unlike equation (\ref{eqn:C_BT_hh_12}), we are not able to simplify the expression by canceling
the power spectrum with its inverse, i.e. equation (\ref{eqn:Pinv_identity}), as $P^{-1}(k)$ does 
not immediately connect to any other $P(k)$. 
Rather, the cancelation would only be achieved by taking the average with another linear field 
$\vpsi^{(1)}(\vk, \eta)$.
We would show that this contribution would be identical to the hard-hard part of the $P_{22}$ term
\begin{eqnarray}
\label{eqn:P22_hdhd}
P^{(22)}_{\tildeLambda\tildeLambda, ab} (k, \eta) 
 &=& 2\biggl \langle   
 \mS^{B^T (\Rmnum{3})}_{\tildeLambda\tildeLambda, a}(\vk, \eta)
  \psi^{(1)}_b (-\vk, \eta)   \biggr\rangle  \nonumber \\
&=&  \frac{1}{2} \times 4 \int^{\eta}_0 ds_1 ~ g_{am}(\eta-s_1)  
 \int_{\tildeLambda\tildeLambda}  ~ d\vq  
 \gamma_{mnc}(\vq, \vk-\vq)   \int_0^{\eta} ds_2 ~g_{d\alpha}(s_1-s_2)  \nonumber \\  
&& \times  \gamma_{\alpha\beta\gamma}(\vq-\vk, -\vq) 
  \biggl [ g_{\gamma\lambda}(s) g_{n\epsilon}(\eta) P^{\ini}_{\lambda\epsilon}(q) \biggr]  
 \biggl [ g_{\beta\delta}(s)  g_{c\tau}(\eta)  P^{\ini}_{\delta\tau}(|\vk-\vq|) \biggr]  
 g_{bd}(\eta-s_1) \nonumber \\
&=& 2  \int_{\tildeLambda\tildeLambda}  d\vq ~  \mF^{(2)}_a (\vk, \vq; \eta) \mF^{(2)}_b( -\vk, -\vq; \eta) 
 P^{\ini} (q) P^{\ini}(|\vk-\vq|) .
\end{eqnarray}
The numerical factor $1/2$ raises from the fact that the time integral in equation 
(\ref{eqn:Shh_BIII}) is limited  by causal constraint $s_2 \le s_1$, which  
could also be seen from the diagram representation in Figure.~ (\ref{fig:P1l_bi}). 
Since there is no difference in deriving this term compared to the one contributing to $P^{13}$,
in this sense, unlike other effective approaches, we interpret this counter term as deterministic
instead of stochastic in our formalism.

\begin{figure*}
	\includegraphics[width=1\textwidth]{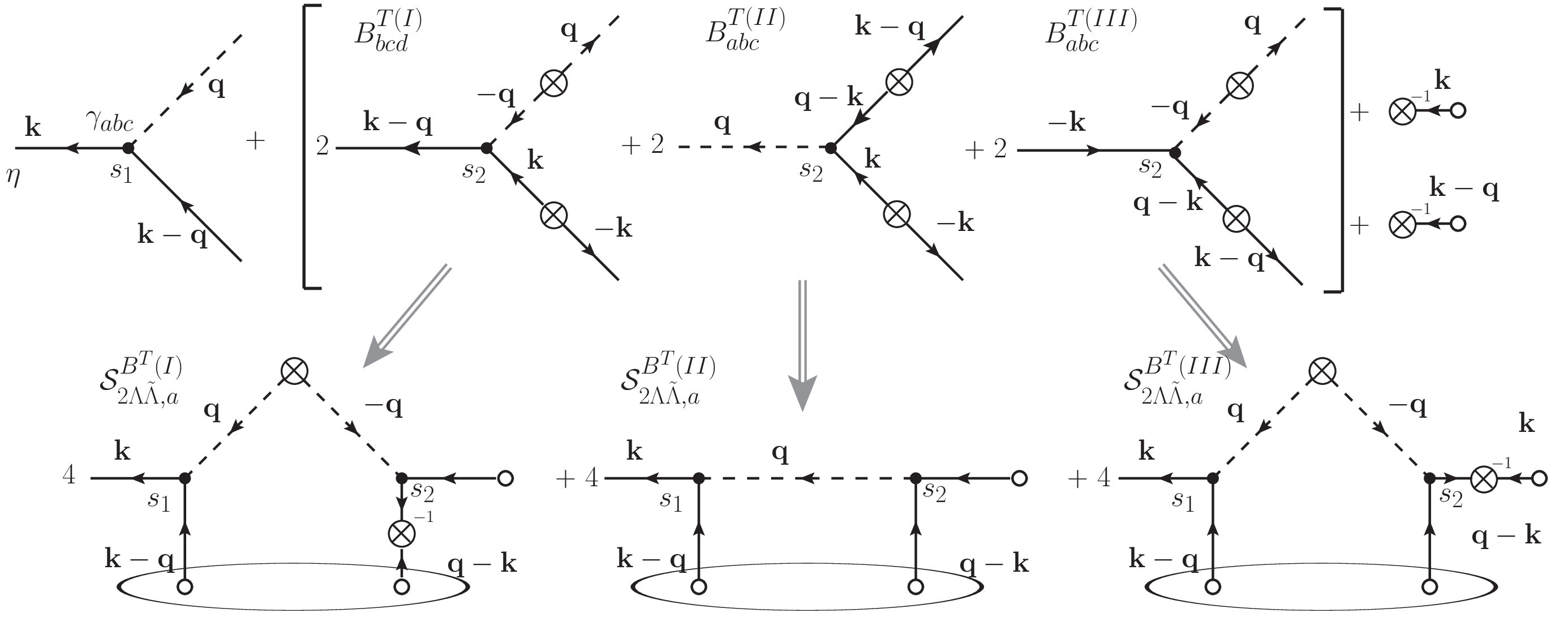}
	\caption{\label{fig:hard_soft_1lp}
		The soft-hard part of the effective terms up to one-loop order. Unlike the hard-hard coupling, 
		these contributions are all {\it non-local} in Fourier space, which means the effective terms for the
		evolution of Fourier mode $\vk$ not only depends on mode $\vk$ but also $\vk-\vq$. 
		For these non-local diagrams, one could see that the first two diagrams at the bottom would form the 
		soft-hard version of $P_{13}(\vk)$, and the third one would form $P_{22}(\vk)$ contribution. 
		The ellipses highlight the pairs that would eventually be connected together when taking
		the ensemble average with another linear field $\vPsi(\vk)$. 
	}
\end{figure*}

\subsection{Soft-hard Coupling}
\label{sec:PT_SHcoupling}
While equation (\ref{eqn:condav_def}) would apply for hard-hard coupling,  the soft-hard modes would be simplified to 
\begin{eqnarray}
	\label{eqn:soft_hard_def}
	\langle \psi_{b,\vk_1} \psi_{c, \vk_2} | \vpsi_{\Lambda} \rangle  =
	\langle \psi_{b,\vk_1} | \vpsi_{\Lambda} \rangle   \psi_{c, \vk_2}. 
\end{eqnarray}

From equation (\ref{eqn:soft_hard_def}) and (\ref{eqn:cond_ave_hs}),  the only non-vanishing  
contribution of soft-hard coupling that is proportional to the bispectrum is a quadratic term
\begin{eqnarray}
\label{eqn:sh_condave}
\langle x_1 | \vY\rangle &=&  \frac{1}{2} \xi^{x_1YY}_{\alpha\beta}   
 \left( \xi^Y \right)^{-1}_{\alpha\lambda}  \left( \xi^Y \right)^{-1}_{\beta\tau} Y_{\lambda}
 Y_{\tau}  .
\end{eqnarray}
Denoting $x_1$ as one component of $\vpsi(\vq)$ and $x_2$  for $\vpsi(\vk-\vq)$ 
where $q>\Lambda$ and $|\vk-\vq| <\Lambda$,  the only relevant bispectrum will
be $B(\vq, \vk-\vq, -\vk)$. Therefore, the effective coefficient in this case could then be 
expressed as
\begin{eqnarray}
\mC^{B^T}_{2\Lambda\tildeLambda,a}(\vk, \eta) &=&
4\times \frac{1}{2} \int_{\Lambda\tildeLambda} d\vq ~ \gamma_{abc}(\vq, \vk-\vq)
B^{T}_{bcd}(\vq,  \vk-\vq, -\vk; \eta)   P^{-1}_{ce}(|\vk-\vq|; \eta)  P^{-1}_{df}(k; \eta)  \nonumber \\
&& \times \psi_{e}(\vq-\vk; \eta) \psi_f(\vk; \eta)   \psi_g(\vk-\vq; \eta). 
\end{eqnarray}
Besides the numerical factor $1/2$ from equation (\ref{eqn:sh_condave}), 
a factor of $2$ raises from the two symmetric Fourier region $\Lambda\tildeLambda$ and 
$\tildeLambda\Lambda$; 
and the other factor of $2$ comes from the fact that equation (\ref{eqn:sh_condave}) have two cross 
contributions when summing over $Y_{\lambda}$ and $Y_{\tau}$.  

Similarly, we could substitute the tree-level bispectrum (equation \ref{eqn:Btree_I}), and obtain
the one-loop order $\mC^{B^T}_{2\Lambda\tildeLambda, a} (\vk, \eta)$. 
For $B^{T(\Rmnum{1})}$, we have
\begin{eqnarray}
\mC^{B^T (\Rmnum{1})}_{2\Lambda\tildeLambda, a} (\vk, \eta) &=& 4  \int_{\Lambda\tildeLambda} d\vq~
\gamma_{abc}(\vq, \vk-\vq) \int_0^{\eta} ds~ g_{d\alpha} (\eta-s)   
\gamma_{\alpha\beta\gamma} (-\vq, \vk) \biggl [ g_{\beta\delta}(s) g_{b\tau}(\eta) 
P_{\delta\tau}^{\ini}(q) \biggr]  \nonumber \\ 
&& \times P^{-1}_{df}(|\vk-\vq|; \eta) \psi_c(\vk-\vq, \eta)
\psi_f(\vq-\vk, \eta)  \psi_{\gamma}(\vk, s)  .
\end{eqnarray}
This equation corresponds to the left-bottom diagram in Figure.~ (\ref{fig:hard_soft_1lp}). 
Clearly, it is {\it non-local} in Fourier space, as the coefficient for Fourier mode $\vk$ also depends on
the mode $\vk-\vq$, where $q>\Lambda$. 
To better understand this term, we have to study the corresponding contribution to the power 
spectrum, i.e. $\left \langle \mS^{B^T (\Rmnum{1})}_{2\Lambda\tildeLambda, a} (\vk) 
\psi^{(1)}_b(-\vk)\right \rangle$. 
From the diagram, the only possible way is to connect $\vk-\vq$ mode with $\vq-\vk$ mode, which
we highlight in the figure.  
And it is clearly that this would produce the soft-hard part of $P_{13}(\vk)$, where the 
integration is over the hard-mode $P(q)$.

For the bispectrum $B^{T (\Rmnum{2})}$, one similarly write down the effective term as
\begin{eqnarray}
\mC^{B^T (\Rmnum{2})}_{2\Lambda\tildeLambda, a} (\vk, \eta) &=&
4  \int_{\Lambda\tildeLambda} d\vq~ \gamma_{abc}(\vq, \vk-\vq) \int_0^{\eta} ds~ 
g_{b\alpha} (\eta-s)    ~ \gamma_{\alpha\beta\gamma} (\vq-\vk, \vk) \psi_c(\vk-\vq; \eta)   \nonumber \\
 && \times~ \psi_{\beta}(\vq-\vk; s)   \psi_{\gamma}(\vk; s), 
\end{eqnarray}
which is presented as the middle-bottom diagram of Figure.~ (\ref{fig:hard_soft_1lp}). 
When taking the ensemble average, $\psi(\vk-\vq)$ will connect with $\psi(\vq-\vk)$, 
therefore this would also contribute to the $P_{13}(k)$, where the integration is over the
soft mode $P(|\vk-\vq|)$. 
Finally, the last term could be expressed as
\begin{eqnarray}
\mC^{B^T (\Rmnum{3})}_{2\Lambda\tildeLambda, a} (\vk, \eta) &=&
4  \int_{\Lambda\tildeLambda} d\vq~ \gamma_{abc}(\vq, \vk-\vq) \int_0^{\eta} ds~ 
g_{b\alpha} (\eta-s)  \gamma_{\alpha\beta\gamma} (\vq-\vk, -\vq)  \biggl[ 
g_{\gamma\lambda}(s) g_{b\epsilon}(\eta) P^{\ini}_{\lambda\epsilon}(q) \biggr] \nonumber \\
&& \times ~ P_{eg}^{-1}(k; \eta) \psi_{\beta}(\vq-\vk; s)  \psi_{c}(\vk-\vq; \eta)  \psi_g(\vk; \eta) .
\end{eqnarray}
From the figure, one could see this contribution will contribute to $P_{22, \Lambda\tildeLambda}(k)$. 
In Table \ref{tab:SPT_comp}, we listed all effective coefficients and their corresponding
contribution in the standard perturbation theory. 
Evidently, this demonstrates that the effective solution is simply a re-organization
of the standard perturbative calculation at one-loop level.
It also indicates that the Fourier non-locality of these effective terms is {\it crucial} to the full 
recovery of the statistical information.

\bibliographystyle{JHEP}
\bibliography{sta_equivalence}

\label{lastpage}

\end{document}